\documentclass[a4paper, 12 pt, twoside, reqno]{amsart}

\usepackage[english]{babel}
\usepackage[utf8]{inputenc}

\usepackage{setspace}
\usepackage[euler-digits,euler-hat-accent]{eulervm}

\usepackage{times}
\usepackage{amsmath}
\usepackage{amsfonts}
\usepackage{amssymb}
\usepackage{amsmath}
\usepackage{amsthm}
\usepackage{graphicx}
\usepackage{array}
\usepackage{color}
\usepackage{mathrsfs}
\usepackage{hyperref}
\usepackage{eucal}
\usepackage{esint} 
\usepackage{tikz}
\usepackage{upgreek}
\usepackage{enumitem}

\usepackage{tikz-cd}

\allowdisplaybreaks

\theoremstyle{plain}
\newtheorem{theorem}{Theorem}[section]
\newtheorem{corollary}[theorem]{Corollary}
\newtheorem{proposition}[theorem]{Proposition}
\newtheorem{lemma}[theorem]{Lemma}

\theoremstyle{definition}
\newtheorem{definition}[theorem]{Definition}

\theoremstyle{remark}
\newtheorem{remark}[theorem]{Remark}

\numberwithin{equation}{section}
\numberwithin{figure}{section}
\numberwithin{table}{section}

\newcommand{\R}{\mathbb{R}}
\newcommand{\N}{\mathbb{N}}
\newcommand{\C}{\mathbb{C}}                           
\newcommand{\Q}{\mathbb{Q}}
\newcommand{\Z}{\mathbb{Z}}

\newcommand{\s}[1]{\CMcal{#1}}
                  
\newcommand{\bb}[1]{\mathscr{#1}}
\newcommand{\rr}[1]{\mathfrak{#1}}
\newcommand{\n}[1]{\mathbb{#1}}

\newcommand{\ketbra}[2]{|#1\rangle\langle#2|}

\newcommand{\expo}[1]{\,\mathrm{e}^{#1}\,}

\newcommand{\dd}{\,\mathrm{d}}
\newcommand{ \ii}{\,\mathrm{i}\,}

\newcommand{\virg}[1]{\lq\lq#1\rq\rq}                
\newcommand{\ie}{\textsl{i.\,e.\,}}
\newcommand{\eg}{\textsl{e.\,g.\,}}
\newcommand{\cf}{\textsl{cf}.\,}
\newcommand{\etc}{\textsl{etc}.\,}

\newlength{\dhatheight}
\newcommand{\doublehat}[1]{%
    \settoheight{\dhatheight}{\ensuremath{\widehat{#1}}}%
    \addtolength{\dhatheight}{-0.3ex}%
    \widehat{\vphantom{\rule{1pt}{\dhatheight}}%
    \smash{\widehat{#1}}}}

\hypersetup{
pdftoolbar=true,        
pdfmenubar=true,        
pdffitwindow=true,     
pdfstartview=true,    
pdftitle={Topology-of-Pure-States},    
pdfauthor={Giuseppe De Nittis and Santiago Rendel},     
breaklinks=true, 
colorlinks=true,       
linkcolor=purple,         
citecolor=teal, 
urlcolor=blue, 
bookmarksopen=true, 
filecolor=magenta,      
}

\begin{document}

\title[Topological description of pure invariant states of the Weyl $C^*$-algebra]{
Topological description of pure invariant states of the Weyl $C^*$-algebra}

\author[G. De~Nittis]{Giuseppe De Nittis}

\address[G. De~Nittis]{Facultad de Matemáticas \& Instituto de Física,
  Pontificia Universidad Católica de Chile,
  Santiago, Chile.}
\email{gidenittis@uc.cl}

\author[S.~G. Rendel]{Santiago G. Rendel}

\address[S.~G. Rendel]{Facultad de Matemáticas \& Instituto de Física,
  Pontificia Universidad Católica de Chile,
  Santiago, Chile.}
\email{sgr@uc.cl}

\vspace{2mm}

\date{\today}

\begin{abstract}
In this work we study the topology of certain families of states of the Weyl $C^*$-algebra with finite degrees of freedom.
We focus on families of pure states characterized by symmetries and a (semi-)regularity condition, and obtain precise topological descriptions through homeomorphisms with other explicit spaces. Of special importance are the families of pure, semi-regular states invariant under either continuous (plane-wave states) or discrete (Bloch-wave states) spatial translations, and the family of states invariant under discrete, mutually commuting spatial and momentum translations (Zak-wave states), all of which we completely characterize. 
\medskip

\end{abstract}

\maketitle

\tableofcontents

\section{Introduction}\label{sec:Intr0}
Pure states of a physical system are the building blocks of more general classes of states. Therefore, they carry basic information about the possible configurations in which the system can exist. Thus, an appropriate understanding of the topology and structure 
of the pure state space should be expected to be relevant for the study and classification of topologically protected phases of the system. 
The works \cite{bhmpqs-24,spiegel-pflaum-25} also investigate this idea, and study topological properties of the pure states of certain classes of $C^*$-algebras. We aim to contribute to this line of work by providing some insight into the topology of the pure states of the Weyl $C^*$-algebra, which is a physical relevant $C^*$-algebra which is not covered by the framework considered in \cite{bhmpqs-24,spiegel-pflaum-25}.

\medskip

Our notion of topological phases, which is related to the one used in \cite{bhmpqs-24}, can be summarized as follows. First, let $\bb{A}$ be a $C^*$-algebra containing the relevant physical observables, and consider a compact, Hausdorff space $X$, which we interpret as a \emph{quantum parameter space}. This space will usually consist of a compactification of the spectrum of a symmetry group $\n{G}$, and its points are interpreted as physically relevant quantum numbers associated with the symmetry. A typical example is provided by the group of \emph{spatial translations} with \emph{momentum} as the associated quantum number.
Let $\s{P}_\bb{A}$ the space of pure states of $\bb{A}$ topologized with the $\ast$-weak topology, and $\s{Q}_\bb{A}\subseteq\s{P}_\bb{A}$  a \emph{relevant} subspace defined by certain properties  like the invariance under the action of a symmetry group $\n{G}$, or given structural  conditions (like  regularity, normality, complete factorization, \etc). 
Any continuous function $X\to \s{Q}_\bb{A}$ is interpreted as a \emph{configuration} (of  type  $\s{Q}_\bb{A}$) of the system, and \emph{topological phases} are the homotopy classes of configurations. 
Of course, to compute the homotopy classes of maps $X\to \s{Q}_\bb{A}$, and thus the topological phases of the system, one must know well the topology and structure of the state space of interest $\s{Q}_\bb{A}$. The goal of this work is to provide a description of the 
relevant subspaces of the pure state space of the Weyl $C^*$-algebra.
On the other hand, the following work \cite{denittis-rendel-25} will cover the description of the associated topological phases and their connection to more general families of states.

\medskip

To be more specific, our starting point is the study of certain families of states of the Weyl $C^*$-algebra $\bb{W}$, for a system with $d\in\N$ degrees of freedom.
The justification for this choice is provided by the fact that $\bb{W}$ is the smallest $C^*$-algebra containing the canonical commutation relations (CCR), and for this reason is the basis for the description of the kinematics of any quantum physical system with $d$ degrees of freedom. Any $C^*$-algebra describing the dynamics of this type of system should contain $\bb{W}$ as a sub-algebra, and in turn any state of this larger algebra must restrict to a state of $\bb{W}$. In summary, the study and classification of the states of $\bb{W}$ is sufficient for the classification of states of any system with a finite number of degrees of freedom. Of course, in this first attempt, we are intentionally forgetting internal degrees of freedom such as the spin of the particles.

\medskip

The basic ideas that topological protected phases are peculiar to homogeneous systems suggests to restrict one's attention to translationally invariant states. By this we mean elements of the state space $\s{S}_\bb{W}$ that are invariant under an action by automorphisms $\tau:\n{G}\to{\rm Aut}(\bb{W})$, where $\n{G}=\R^d$ for continuous translations and $\n{G}=\Gamma\simeq\Z^d$ for discrete translations. In both cases, states that are $\n{G}$-invariant cease to be regular (\ref{lemma_01} - \ref{lemma_01_01}). This fact has the physical meaning of the loss of the position as a \virg{good} quantum number, as a consequence of the delocalization arising from the invariance under translations of the associated expectation values. Nevertheless, one can still insist in the identification of the momentum as a \virg{good} quantum number, and this translates to imposing a semi-regularity condition on the states of interest. Therefore, there are physical reasons to consider interesting the study and classification of $\n{G}$-invariant semi-regular states. In the special case $\n{G}=\n{R}^d$, we will spend some effort also in the study of irregular states (see Section \ref{sec:classif_sts_tras}) with the purpose to postpone an appropriate analysis of these statistics to a future work. 

\medskip

We start by analysing the set $\s{S}_{\bb{W}}^\tau$ of states invariant with respect to continuous translations ($\n{R}^d$-invariant states). In Proposition \ref{cor:trasnl_st} we provide a complete characterization of these states, including a description of the subset of pure translation-invariant states $\s{P}_\bb{W}^\tau$. Our result generalizes the existing literature \cite[Proposition 3.5]{beaume-manuceau-pellet-sirugue-74} and  \cite[Proposition 2.1]{strocchi-15} in two directions. First of all, we consider the full set $\s{P}_\bb{W}^\tau$, not restricting to the subset $\s{P}_\bb{W}^{\tau,\beta}$ of semi-regular translation invariant pure states. Secondly, in Proposition \ref{prop:homeo_t_inv} we describe $\s{P}_\bb{W}^\tau$ and  $\s{P}_\bb{W}^{\tau,\beta}$ as topological spaces (with the induced $\ast$-weak topology), providing in particular that the latter space is path-connected, \ie $\pi_0(\s{P}_\bb{W}^{\tau,\beta})=\{\ast\}$. 
It is worth mentioning at this point that the path connectedness of this class of states strongly depends on the assumption of semi-regularity. In fact, as discussed in Remark \ref{rk:irreg_phas}, there are irregular states in $\s{P}_\bb{W}^\tau$ which cannot be continuously deformed into semi-regular states. However, the physical meaning of these irregular phases is beyond the scope of this work, and we will keep considering the condition of semi-regularity as an important ingredient for our study. 

\medskip

We then turn our attention to the set of states invariant under discrete translations by some lattice $\Gamma\simeq\Z^d$, denoted $\s{S}_{\bb{W}}^\Gamma$. The reduced symmetries in these states limit the scope of what one can achieve in terms of topological descriptions of subfamilies, partly as a consequence of the $C^*$-algebra of observables $\bb{V}_\Gamma$ fixed under these translations not being commutative. Despite this, Theorem \ref{thm:homeo_gamma_inv} shows that the space of pure, $\Gamma$-invariant, semi-regular states $\s{P}_\bb{W}^{\Gamma,\beta}$ has (up to a homeomorphism) the structure of a  (trivial) fiber bundle. The base space of this bundle is the \emph{Brillouin zone} associated with the lattice $\Gamma$, and the fibers are Hilbert Grassmannians of rank 1 (the set of one-dimensional orthogonal projectors of a separable Hilbert space equipped with the weak operator topology). 
It turns out that $\s{P}_\bb{W}^{\Gamma,\beta}$ is also path-connected.

\medskip

Finally, we briefly study Zak states, corresponding to states invariant under discrete spatial translations by some lattice $\Gamma$, and discrete momentum translations by the dual lattice $\Gamma'$ of $\Gamma$. These are $\Gamma$-invariant states that are not semi-regular, and thus represent a class of $\Gamma$-invariant states disjoint to the one discussed above. Through the additional structure provided by the extra symmetry, we recover some results analogous to those obtained for $\s{P}_\bb{W}^{\tau,\beta}$. In particular, we show that the space of pure Zak states is homeomorphic to a $2d$-dimensional torus.

 \medskip
 \noindent
{\bf Structure of the paper.}
In {\bf Section~\ref{sec:top_class}} we will introduce basic concepts regarding the topology of the state space and the notion of path-equivalence of states.
{\bf Section~\ref{sec:obs_alg_sts}}
describes the basic facts about  the Weyl $C^*$-algebra and its symmetries. 
{\bf Section~\ref{sec:classif_sts_tras}} deals with the topology of pure states invariant under continuous translations and
{\bf Section~\ref{sec:classif_sts_latt}} is concerned with the topology of pure semi-regular states invariant under discrete translations.
In {\bf Section~\ref{app-zac}}  Zak states are presented as an additional class of interesting physical states.
 {\bf Appendix~\ref{sec:topol}} contains some fundamental results about the  theory of fiber bundles and Hilbert Grassmannians.
Finally, {\bf Appendix~\ref{appendix:Bohr}} contains a succinct introduction to the Bohr compactification, especially for the case of $\R^d$.

 \medskip
 
 \noindent
{\bf Acknowledgements.}
GD is supported by the grant \emph{Fondecyt Regular - 1230032}. This research was partially supported by the University of Warsaw Thematic Research Programme ``Quantum Symmetries".


\section{Topology of the state space}
\label{sec:top_class}

An in-depth presentation of the concept of states of $C^*$-algebras and their $\ast$-weak topology can be found in \cite{bratteli-robinson-87}.

\subsection{Basic concepts}
Let $\bb{A}$ be a unital $C^*$-algebra. A \emph{state} $\omega$ on $\bb{A}$ is a positive, linear and normalized functional.
The state space of $\bb{A}$ will be denoted with
 $\s{S}_\bb{A}$. A state is called \emph{pure} if it cannot be decomposed as a convex combination of two non-zero distinct states.
 The symbol  $\s{P}_\bb{A}$ will be used for
the subset of pure states of $\bb{A}$. The space $\s{S}_\bb{A}$ can be endowed with different topologies. For many reasons it is useful to assign it the $\ast$-weak topology, that is, the topology  of point-wise convergence of nets on  $\s{S}_\bb{A}$.
A basis of this topology is provided by the family of neighbourhoods of any $\omega\in \s{S}_\bb{A}$ defined by
\[
\mathtt{U}_{\{a_1,\ldots,a_n\},\varepsilon}(\omega)\;:=\;\left\{\omega'\in\s{S}_\bb{A}\;|\; |\omega(a_i)-\omega'(a_1)| < \varepsilon\;, \quad i=1,\ldots,n\right\}
\]
and indexed by $\varepsilon>0$ and finite sets of elements $\{a_1,\ldots,a_n\}\subset\bb{A}$. It is well known that $\s{S}_\bb{A}$ is a compact Hausdorff space with respect to this topology. Moreover it is a convex space whose set of extremal points coincides with $\s{P}_\bb{A}$, and the full space  $\s{S}_\bb{A}$ can be obtained as the closure of  the convex envelope of $\s{P}_\bb{A}$ \cite[Theorem 2.3.15]{bratteli-robinson-87}.

\begin{remark}\label{rk:state_metriz_alg_sep}
The $\ast$-weak topology on $\s{S}_\bb{A}$ is metrizable when $\bb{A}$ is separable. However, this is not the case for the Weyl $C^*$-algebra, which is the main algebra considered in this work.
 \hfill $\blacktriangleleft$
\end{remark}

Given any subset $\s{K}\subseteq\s{S}_\bb{A}$ we will consider it as a topological space endowed with the subspace (or induced) topology. Thus $\s{K}$ is automatically Hausdorff. When $\s{K}$ is closed in $\s{S}_\bb{A}$, then $\s{K}$ is again a compact space with respect to the induced topology. However, this is not always the case of interest. For instance, the subset $\s{P}_\bb{A}$ is a Baire space (with respect to the the subspace topology) \cite[Corollary 5.2.8]{saito-wright-15} but in general it may fail to be closed in $\s{S}_\bb{A}$.

\subsection{Equivalence of states}

A continuous function $X\to \s{Q}_\bb{A}\subseteq \s{P}_\bb{A}$, called a \emph{configuration}, can essentially be conceived as a continuous collection of pure states in $\s{Q}_\bb{A}$, they can give rise to more general classes of mixed states obtained by an integral of this collection. As such, one may wonder whether \emph{topological phases}, \ie, homotopy classes of configurations, can be identified with equivalence classes of states under an appropriate equivalence relation. This is what we would call a \emph{topological classification of states}, and while we will not study this in depth in the current work, the idea provides a few directives of importance.

For an eventual classification of states, we will need a notion that describes when two states are equivalent. A natural idea is to say that two states are equivalent if they can be continuously deformed into each other. However, the concept of continuity depends on the topology. For this reason, we will always assume that $\s{S}_\bb{A}$ is topologized with the $\ast$-weak topology, and any subset  $\s{K}\subseteq\s{S}_\bb{A}$ with the induced topology

\begin{definition}[Path-equivalence of states]
Given a subset $\s{K}\subseteq\s{S}_\bb{A}$ and two states $\omega_0,\omega_1\in \s{K}$ we will say that the two states are \emph{path-equivalent (in $\s{K}$)} if and only if there exists a continuous map $[0,1]\ni t\mapsto \omega_t\in\s{K}$ joining $\omega_0$ with $\omega_1$. \end{definition}

\medskip

This notion of equivalence can be argued to be too weak for general states, but it provides an important ingredient for a more appropriate definition. In fact, if a subset $\s{Q}_\bb{A}\subseteq \s{P}_\bb{A}$ is not path-connected, then for any compact, Hausdorff space $X$, the set $[X,\s{Q}_\bb{A}]$ of homotopy classes of functions $X\to\s{Q}_\bb{A}$ is non-trivial, \ie, it has more than one element. This would mean that these states manifest distinct phases independently of the quantum parameter space in question.

\medskip

The path-equivalence class of a state $\omega\in \s{K}$ inside $\s{K}$ will be denoted by $[\omega]_{\s{K}}$, or simply by $[\omega]$ when the ambient space is clearly specified by the context. 
Additionally, the symbol ${\pi_0}(\s{K})$ will be used for the set of path-equivalence classes of states in $\s{K}$. In other words ${\pi_0}(\s{K})$ counts the path components of $\s{K}$.

\medskip

Observe that, since the full state space $\s{S}_\bb{A}$ of any $C^*$-algebra $\bb{A}$ is a convex space, it turns out to be automatically path-connected. Therefore, one immediately gets ${\pi_0}(\s{S}_\bb{A})=\{[\omega_\ast]\}$, namely the set of path-equivalence classes of states reduces to a singleton which can be represented by any reference state $\omega_\ast$. 

\medskip

Before ending this section, we find it prudent to introduce a stronger notion of equivalence, which has already been employed and studied in the literature dealing with topological phases in interacting systems.

\begin{definition}[Automorphic equivalence of states]
    Fix a subset $\s{K}\subseteq\s{S}_\bb{A}$ and two states $\omega_0,\omega_1\in \s{K}$. We will say that the two states are \emph{automorphically equivalent (inside $\s{K}$)} if and only if there exists a strongly continuous map $[0,1]\ni t\mapsto \lambda_t\in{\rm Aut}(\bb{A})$, such that $t\mapsto \omega_t := \omega_0 \circ \lambda_t$ is a continuous path in $\s{K}$ joining $\omega_0$ and $\omega_1$.
\end{definition}

We will not often use this definition, as it is too strong in the context of the Weyl $C^*$-algebra $\bb{W}$. This is due to the fact that, as the \virg{smallest} $C^*$-algebra containing the CCR, $\bb{W}$ contains few automorphisms in a precise sense. For example, in the Schr\"odinger representation, time evolution automorphisms given by a wide range of Hamiltonians do not preserve the (represented) Weyl $C^*$-algebra \cite{fannes-verbeure-74}.
Despite this, we will still show some families of pure states that are all both equivalent and automorphically equivalent to each other.

\section{Observable algebra and invariant states}
\label{sec:obs_alg_sts}
This section is devoted to the abstract construction of the Weyl $C^*$-algebra
and the presentation of its main properties. We also describe the structure of 
its translationally invariant states and the related GNS representations.
The material presented here is part of the standard literature concerning the mathematical foundations of quantum mechanics. For the benefit of the reader, we will refer mainly to the papers \cite{manuceau-sirugue-testard-verbeure-73,slawny-72,beaume-manuceau-pellet-sirugue-74} and to the monographs \cite{bratteli-robinson-97,petz-90,strocchi-08}.

\subsection{The Weyl \texorpdfstring{$C^*$}{TEXT}-algebra}\label{sec_Q_W}
The ($d$-dimensional) \textit{Weyl algebra}  (also known as \textit{CCR algebra}) is the algebra generated by the family of non-zero elements $\{u_\alpha, v_\beta\;|\;\alpha,\beta\in\R^d\}$, that satisfy the product laws 
\begin{equation}\label{weylrelmix0}
\begin{aligned}
    u_\alpha v_\beta\; =\; \expo{\ii \alpha\cdot \beta}\;v_\beta u_\alpha \;,\quad
   u_\alpha u_{\alpha'} \; =\; u_{\alpha + \alpha'}\;,\quad     v_\beta v_{\beta'}\; =\; v_{\beta + \beta'}\;
    \end{aligned}
\end{equation}
where $\alpha\cdot \beta$ denotes the usual  Euclidean scalar product in $\R^d$.
These relations are called the \textit{canonical commutation relations (CCR) in Weyl form} (or {Weyl relations} for short) and $d$ expresses the \emph{degrees of freedom} of the system. 
 Because of \eqref{weylrelmix0}, one has that all monomials of products of the generators reduce to elements of the form $\expo{\ii c} u_\alpha v_\beta$, for some $\alpha,\beta\in\R^d$ and $c\in\R$.
Therefore,  every element of the algebra is a linear combination of elements of this form. It can be endowed with the $\ast$-involution given by
\begin{equation}\label{weylrelmix1}
\begin{aligned}
    u_\alpha^* \; =\;  u_{-\alpha} \;,\qquad
    v_\beta^* \; =\; v_{-\beta}\;.
    \end{aligned}
\end{equation}
From the relations \eqref{weylrelmix0} and the assumption that the $u_\alpha$ and $v_\beta$ are non-zero one concludes that $u_0={\bf 1}={v_0}$ (the identity of the algebra) and from \eqref{weylrelmix1} one infers that $u_\alpha^* = u_\alpha^{-1}$ and $v_\beta^* = v_\beta^{-1}$ (unitary generators). The Weyl $\ast$-algebra, constructed as the formal complex linear combinations of the monomials $u_\alpha v_\beta$, will be denoted as $\bb{W}_0$.
\begin{remark}[Presentation by the symplectic structure]\label{rk:symp_str}
The Weyl relations are usually presented by using the standard symplectic structure on the cotangent bundle (phase space) of $\R^d$.  Let us use the standard identification $T^*\R^d\simeq\R^d\times\R^d$, which allows us to adopt the convention $z=(\alpha,\beta)$ for a generic point of $T^*\R^d$. Let $\sigma:T^*\R^d\times T^*\R^d \to\R$ be the \emph{symplectic product}
\[
\sigma(z,z')\;:=\;\frac{1}{2}(\alpha\cdot\beta'-\alpha'\cdot\beta)\;,\qquad z,z'\in T^*\R^d\;.
\]
Then the $\ast$-algebra $\bb{W}_0$ can be equivalently defined as the algebra generated by the family of non-zero elements $\{w_z\;|\;z\in T^*\R^d\}$ which satisfy the relations
\[
w_zw_{z'}\;=\;\expo{\ii\sigma(z,z')}w_{z+z'}\;,\qquad w_z^*\;=\;w_{-z}\;.
\]
The equivalence between the two definitions is given by the relations of the generators: $w_{(\alpha,\beta)}:=\expo{-\frac{\ii}{2}\alpha\cdot\beta}u_\alpha v_\beta$. 
 \hfill $\blacktriangleleft$
\end{remark}

There are many good reasons to complete a $\ast$-algebra to a $C^*$-algebra.
A standard procedure is to introduce the \emph{minimal regular (semi-)norm}, which in our case is given by
\begin{equation}\label{eq:u_norm}
\|a\|\;:=\;\sup_{\rho}\left\{\|\rho(a)\|_{\bb{B}(\s{H}_{\rho})}\right\}\;,\qquad a\in\bb{W}_0\;.
\end{equation}
where the supremum is taken over all possible $\ast$-representations $\rho:\bb{W}_0\to\bb{B}(\s{H}_{\rho})$ of $\bb{W}_0$ as bounded operators on some Hilbert space $\s{H}_{\rho}$ \cite[Section 2.7]{dixmier-77}. 
\begin{remark}[Schr\"odinger representation]\label{rk:ex_rep}
It is worth mentioning that definition \eqref{eq:u_norm} is well posed since there are concrete representations of $\bb{W}_0$. 
The most relevant is the Schr\"odinger representation
 $\rho_S$ on the  the (separable) Hilbert space $\s{H}_S:=L^2(\R^d)$ defined by
 \[
[\rho_S(u_\alpha)\psi](x)\;:=\;\expo{\ii \alpha\cdot x}\psi(x)\;,\qquad [\rho_S(v_\beta)\psi](x)\;:=\;\psi(x-\beta)\;,
\]
for every $\psi\in\s{H}_S$. This representation is irreducible and corresponds to the standard formulation of
 quantum mechanics.
\hfill $\blacktriangleleft$
\end{remark}

\medskip

It turns out that $\|\cdot\|$ is indeed a norm \cite[Lemma 3.1]{manuceau-sirugue-testard-verbeure-73}, and the completion of $\bb{W}_0$ with respect to this norm, denoted with
\[
\bb{W}\;:=\;\overline{\bb{W}_0}^{\;\|\cdot\|}
\]
will be called the \emph{(abstract) Weyl $C^*$-algebra}. The most structurally important aspect that characterizes $\bb{W}$ is expressed in the following classical result.
\begin{theorem}\label{th:mai}
The $C^*$-algebra $\bb{W}$  is the unique $C^*$-completion of $\bb{W}_0$,  up to isomorphism. It is simple and non-separable.
\end{theorem}
\medskip

The first proof of this result was given in \cite{manuceau-sirugue-testard-verbeure-73}. However, useful references are also \cite[Theorem 2.1]{petz-90} and \cite[Theorem 5.2.8]{bratteli-robinson-97}. The unicity of $\bb{W}$ is a consequence of the fact that the norm \eqref{eq:u_norm} is indeed the unique $C^*$-norm that provides a $C^*$-completion of $\bb{W}_0$ \cite[Corollary 4.23]{manuceau-sirugue-testard-verbeure-73}. The simplicity (absence of non-trivial ideals) of $\bb{W}$ \cite[Corollary 4.24]{manuceau-sirugue-testard-verbeure-73} implies that any $C^*$-representation of $\bb{W}$ is indeed \emph{faithful} (\ie an isomorphism). Two facts that are consequences of Theorem \ref{th:mai} will be of repeated use throughout this work and, for this reason, we will present them in a precise form.
\begin{proposition}\label{pro_constr_rep}
Let $\s{H}$ be a Hilbert space and $\{U(\alpha), V(\beta)\;|\;\alpha,\beta\in\R^d\}$ a family of unitary operators such that $U(\alpha)V(\beta)= \expo{\ii \alpha\cdot \beta}V(\beta)U(\alpha)$ for every $\alpha,\beta\in\R^d$. Then the mapping $\rho:u_\alpha\mapsto U(\alpha)$, $\rho:v_\beta\mapsto V(\beta)$ extends to a faithful representation $\rho:\bb{W}\to\bb{B}(\s{H})$.
\end{proposition}

\medskip

Given a representation $\rho$ of $\bb{W}$ on the Hilbert space $\s{H}$, we denote by $\rr{M}_\rho:=\rho(\bb{W})''\subseteq \bb{B}(\s{H})$ the \emph{enveloping} von Neumann algebra generated by the bicommutant of $\rho(\bb{W})$. When $\rho$ is irreducible one immediately obtains that $\rr{M}_\rho=\bb{B}(\s{H})$ \cite[Proposition 2.3.8]{bratteli-robinson-87}. Although all the realizations of the Weyl $C^*$-algebra are mutually isomorphic, the various {enveloping} von Neumann algebras can be very distinct due to the specific nature of the representing Hilbert space. 
\begin{proposition}\label{pro:stat}
Any positive linear funcional on $\bb{W}_0$
 extends to a unique positive linear functional on $\bb{W}$.
\end{proposition}

\medskip

This result is a direct consequence of \cite[Proposition 2.17]{manuceau-sirugue-testard-verbeure-73} and the definition of the minimal regular norm \eqref{eq:u_norm}.
Let us denote with $\s{S}_\bb{W}$ the \emph{state space} of $\bb{W}$, \ie the set of normalized positive linear functionals over  $\bb{W}$. The subspace of \emph{pure} states (the extremal points of $\s{S}_\bb{W}$) will be denoted with $\s{P}_\bb{W}$. The relevance of Proposition \ref{pro:stat} lies in the fact that, to define an element $\omega\in \s{S}_\bb{W}$, it is sufficient to specify the values $\omega(u_\alpha v_\beta)$ for all $\alpha,\beta\in\R^d$, in such a way that the extension by linearity on $\bb{W}_0$ produces a positive functional. The normalization condition forces $\omega(u_0 v_0)=\omega({\bf 1})=1$. Using this recipe, it is possible to construct a relevant state on $\bb{W}$ defined by the conditions
\begin{equation}\label{eq:tr_st}
\rr{t}(u_\alpha v_\beta)\;:=\;\delta_{\alpha,0}\delta_{\beta,0}\;,\qquad \alpha,\beta\in\R^d\;.
\end{equation}
The associated state $\rr{t}$ is called the \emph{(canonical) tracial}  state of $\bb{W}$, since  it can be shown that  $\rr{t}(ab)=\rr{t}(ba)$ for all $a,b\in \bb{W}$. 
This special state can be used to prove a couple of important properties of $\bb{W}$. First of all, one can deduce the inequality
\begin{equation}\label{eq:norm_ineq}
\|\lambda u_\alpha v_\beta-\lambda' u_{\alpha'} v_{\beta'}\|\;\geqslant\;\sqrt{|\lambda|^2+|\lambda'|^2} \;,\quad\text{if}\;\; \alpha\neq\alpha'\;\;\text{or}\;\;\beta\neq\beta'
\end{equation}
with $\lambda,\lambda'\in\C$ \cite[Proposition 2.2]{petz-90}.
The second result concerns an explicit series representation for a generic element $a\in \bb{W}$. As proved in \cite[Lemma II.4]{fannes-verbeure-74}
one has that 
\begin{equation}\label{eq:exp_l2}
a\;:=\;\sum_{(\alpha,\beta)\in\Sigma_{g_a}}g_a(\alpha,\beta)\;u_\alpha v_\beta
\end{equation}
where $g_a\in\ell^2(\R^{2d})$ is a function with countable support $\Sigma_{g_a}$ and $\ell^2$-summability property defined by $g_a(\alpha,\beta):=\rr{t}(u_\alpha^* v_\beta^* a)$ for every $(\alpha,\beta)\in\Sigma_{g_a}$, and zero otherwise. The convergence of \eqref{eq:exp_l2} is meant in the $\ell^2$-sense.

\begin{definition}[Regular and semi-regular states]\label{def:r_s}
A state $\omega\in \s{S}_\bb{W}$ is called \emph{regular} if the map
\[
\R^2\;\ni\; (t,s)\;\longmapsto\;\omega(u_{t\alpha}v_{s\beta})\;\in\C
\]
is continuous for every $\alpha,\beta\in\R^d$. We will say that $\omega$ is semi-regular if it satisfies this condition on only in one of the two parameters $t$ or $s$.
\end{definition}

\medskip

It turns out that for $\omega$  regular in the associated GNS representation $(\pi_\omega, \s{H}_\omega, \psi_\omega)$ the unitary groups $\alpha\mapsto \pi_\omega(u_\alpha)$ and $\beta\mapsto \pi_\omega(v_\beta)$
are strongly continuous and can be differentiated along their parameters. A similar observation holds true for a semi-regular state with respect to the regular parameter.
An example of a regular state is given by the \emph{Fock state} $\omega_F$ defined by the relations
\begin{equation}\label{eq:F_S}
\omega_F(u_{\alpha}v_{\beta})\;=\;\expo{\frac{\ii}{2}\alpha\cdot\beta}\;\expo{-\frac{|\alpha|^2+|\beta|^2}{4}}\;.
\end{equation}
It is pure since the induced GNS representation, which coincides with the Schr\"odinger representation, is irreducible.

\medskip

On the other hand, the tracial state described in \eqref{eq:tr_st}, or the \emph{Zak states} described in Appendix \ref{app-zac}, provide examples of \emph{irregular states}, namely states which are discontinuous in both the parameters  $\alpha$ and $\beta$.

\medskip

Let us end this section by introducing one last ingredient that will be used later on. For every $t\in\R$, consider the mapping defined on the monomials by
\begin{equation}\label{eq:free_dyn}
\Phi_t(u_\alpha v_\beta)\;=\; \expo{\ii\frac{t}{2}|\alpha|^2}u_\alpha v_{\beta- t\alpha}\;.
\end{equation}
 This extends by linearity to a $\ast$-automorphism of $\bb{W}$. Moreover, the mapping $\R\ni t\mapsto \Phi_t\in{\rm Aut}(\bb{W})$ provides a representation of $\R$ as $\ast$-automorphisms of $\bb{W}$, and will be called the \emph{free dynamics}. In view of \eqref{eq:norm_ineq} it turns out that $t\mapsto \Phi_t$ is not strongly continuous, and thus it does not provide a $C_0$-group. Consequently, there is no natural infinitesimal generator associated to  $\bb{W}$ which realizes the dynamics $\Phi_t$. Let us also point out that the free dynamics does not preserve the Fock state. Indeed, a direct check shows that 
$\omega_F\circ \Phi_t \neq\omega_F$ if $t\neq0$.

\begin{remark}[Group algebra structure]\label{rk:symp_str-K}
The presentation described in Remark \ref{rk:symp_str} allows the identification of the Weyl $C^*$-algebra as a twisted group algebra. More precisely, let $\R^{2d}_d$ be the additive group $T^*\R^d\simeq\R^{2d}$ endowed with the discrete topology. 
Consider the map $\sigma$ defined in Remark \ref{rk:symp_str} as a 
 \emph{group 2-cocycle}. Then one has the isomorphism of $C^*$-algebras
 \[
 \bb{W}\;\simeq\;C^*(\R^{2d}_d,\sigma)
\]
where on the right-hand side one has the twisted group $C^*$-algebra of  $\R^{2d}_d$ \cite[Corollary 3-12]{binz-honegger-rieckers-04}. Isomorphisms of this type are often useful for computations in $K$-theory (\eg  the Baum-Connes conjecture). 
 \hfill $\blacktriangleleft$
\end{remark}

\subsection{Translation-invariant states}\label{sec_tr_inv_st}
The Weyl $C^*$-algebra can be endowed with several groups of symmetries, and in turn one is led to study the structure of the invariant states. A quite general discussion about this problem is given in \cite{beaume-manuceau-pellet-sirugue-74}. We will focus mainly on the symmetries given by the group of \emph{space translations}. This is done by endowing   $\bb{W}$  with the action of
 $\R^d$  by the automorphisms
$\R^d\ni\lambda\mapsto \tau_\lambda\in{\rm Aut}(\bb{W})$   defined by
\[
\tau_\lambda(a)\;:=\;v_\lambda av_\lambda^*\;,\qquad a\in\bb{W}\;.
\]
\begin{proposition}\label{prop:transl_str_cont}
The map $\R^d\ni\lambda\mapsto \tau_\lambda\in{\rm Aut}(\bb{W})$ is strongly continuous.
\end{proposition}
\proof
We first show that $\tau_\lambda$ is norm preserving on $\bb{W}$. Let $a\in\bb W$ and $\lambda\in\R^d$. Take any representation $\pi$ of $\bb W$ on some Hilbert space $\s H$. Then, 
\[
\|\pi(\tau_\lambda(a))\|_{\bb{B}(\s{H})}\;=\;\|\pi(v_\lambda)\pi(a)\pi(v_\lambda)^*\|_{\bb{B}(\s{H})}\;=\;\|\pi(a)\|_{\bb{B}(\s{H})}
\]
since $\pi(v_\lambda)$ are unitary operators. Then, since $\pi$ is faithful, $\|\tau_\lambda(a)\| = \|a\|$ for the abstract algebra as well.
Now, let $a\in \bb W$ and $b\in \bb{W}_0$ such that $\|a-b\|<\epsilon/3$. Note that we may write $b$ as a finite sum $b := \sum_{j=1}^N c_j u_{\alpha_j} v_{\beta_j}$, and thus
$$\|\tau_\lambda(b)-b\|\; =\; 
\left\|\sum_{j=1}^N c_j (e^{-\ii \alpha_j\cdot \lambda}-1) u_{\alpha_j} v_{\beta_j}\right\| \;\leq\; \sum_{j=1}^N |c_j| |e^{-\ii \alpha_j\cdot \lambda}-1|\; \xrightarrow{\lambda\to 0}\;0\;.$$
Then, there exists $\delta>0$ such that $|\lambda|\leq\delta$ implies $\|\tau_\lambda(b)-b\|\leq\epsilon/3$, and we have
\begin{align*}
    \|\tau_{\lambda}(a)-a\| & \;\leq\; \|\tau_{\lambda}(b)-b\| +\|\tau_{\lambda}(a-b)-(a-b)\| \\
    &\;\leq\; \|\tau_{\lambda}(b)-b\| + \frac{2}{3}\epsilon \;\leq\; \epsilon\;.
\end{align*}
Thus $\lim_{\lambda\to0} \|\tau_{\lambda}(a)-a\| = 0$, and $\lambda\mapsto \tau_\lambda$ is strongly continuous. \qed

\medskip
Let 
\[
{\rm Inv}_\tau(\bb{W})\;:=\;\{a\in\bb{W}\;|\;  \tau_\lambda(a)=a\;,\;\;\forall\; \lambda\in\R^d\}
\]
be the space of elements left invariant by translations, and consider the commutative $C^*$-subalgebra $\bb{V}\subset \bb{W}$ generated solely by the $v_\beta$'s, \ie
\[
\bb{V}\;:=\;C^*(\{v_\beta\;|\;\beta\in\R^d\})\;.
\]

\begin{remark}\label{rk:ab_bhor_comp}
In the following, we will make use of
the characterization of the abelian  $C^*$-algebra $\bb{V}$ as an \virg{algebra 
of functions}. More specifically, it can be proven that  
\[
\bb{V}\;\simeq\;AP(\R^d)\;\simeq\; C(\rr{b}(\n{R}^d))
\]
where $AP(\R^d)$ denotes the algebra of  \emph{almost periodic functions}
 on $\R^d$  and $\rr{b}(\n{R}^d)$ is the \emph{Bohr compactification} of $\R^d$ (see Appendix \ref{appendix:Bohr}).
  The first identification is provided by the mapping $v_\beta\mapsto \xi_\beta$
 where $\xi_\beta(x):=\expo{\ii \beta\cdot x}$.
  The identification $AP(\R^d)\simeq C(\rr{b}(\n{R}^d))$ 
is described in Appendix  \ref{sec_AP}. 
 The identification $\bb{V}\simeq C(\rr{b}(\n{R}^d))$ corresponds to the \emph{Gelfand isomorphism}
and $\rr{b}(\n{R}^d)$ provides a model for the \emph{Gelfand spectrum} of $\bb{V}$. We will use the same symbol $\xi_\beta$ for the   
Gelfand transform of $v_\beta$.
\hfill $\blacktriangleleft$
\end{remark}

\medskip

To carry out a detailed analysis of the structure of ${\rm Inv}_\tau(\bb{W})$, we need the definition of the \emph{ergodic mean} of an element $a\in \bb{W}$. This is given by
\begin{equation}\label{eq:er_mean}
\langle a\rangle\;:=\;\lim_{L\to+\infty}\frac{1}{|\Lambda_L|}\int_{\Lambda_L}\dd \lambda\; \tau_\lambda(a)
\end{equation} 
where $\Lambda_L:=[-L,+L]^d$ is the $d$-dimensional cube  of size $2L$ and   
$|\Lambda_L|=(2L)^d$ is its volume with respect to the Lebesgue (Haar) measure $\dd\lambda$ of $\R^d$. The integrals on the right-hand side are meant in the Bochner sense (see \cite[Appendix B]{williams-07} for more details) and the convergence in norm of the sequence of integrals is guaranteed by \cite[Theorem 20]{warren-21}.
Evidently $\langle a\rangle\in{\rm Inv}_\tau(\bb{W})$ by the invariance of the measure and in turn \eqref{eq:er_mean} provides a linear map $\langle \cdot\rangle:\bb{W}\to {\rm Inv}_\tau(\bb{W})$
which results to be norm continuous in view of the inequality $\|\langle a\rangle\|\leqslant \|a\|$.

\begin{proposition}\label{prop_inv_tau_charac}
It holds true that ${\rm Inv}_\tau(\bb{W})= \bb{V}$.
\end{proposition}
\proof
The inclusion $\bb{V}\subseteq{\rm Inv}_\tau(\bb{W})$ follows from the commutativity of  $\bb{V}$.
For the opposite inclusion, let us observe that $\langle a\rangle=a$ for every $a\in {\rm Inv}_\tau(\bb{W})$. This shows that every element of ${\rm Inv}_\tau(\bb{W})$ is of the form $\langle a\rangle$ for some $a\in \bb{W}$. A direct computation shows that
\[
\langle u_\alpha v_\beta\rangle\;=\;\left(\lim_{L\to+\infty}\frac{1}{|\Lambda_L|}\int_{\Lambda_L}\dd \lambda\; \expo{-\ii\alpha\cdot\lambda}\right)u_\alpha v_\beta\;=\;\delta_{\alpha,0} v_\beta\;\in\; \bb{V}\;.
\]
Therefore $\langle \cdot\rangle:\bb{W}_0\to \bb{V}$ by linearity,
and then from the continuity of $\langle \cdot\rangle$ one infers ${\rm Inv}_\tau(\bb{W})\subseteq\bb{V}$. 
\qed

\medskip

States invariant under the group of translations will play an important role in this work.

\begin{definition}[Translationally invariant states]
A state $\omega\in\s{S}_\bb{W}$ is \textit{translationally invariant} if 
$\omega\circ \tau_\lambda   = \omega$ for all $\lambda\in\R^d$. The set of 
translationally invariant states will be denoted with $\s{S}_{\bb{W}}^\tau$.
\end{definition}

\medskip

Let us start by highlighting an important property of translationally invariant states. The next result is a special case of  \cite[Lemma 2.13]{beaume-manuceau-pellet-sirugue-74}.
\begin{lemma}\label{lemma_01}
Let $\omega\in\s{S}_{\bb{W}}^\tau$ be a translationally invariant state. Then
\begin{equation}
    \label{eccond2}
    \begin{aligned}
    \omega(u_\alpha v_\beta)\;&=\;0,\;&&\forall\;\alpha\in\R^d\setminus\{0\}\;,\;\;\forall\;\beta\in\R^d\;.\\ 
    \end{aligned}
   \end{equation}
As a consequence, the map $t\mapsto \omega(u_{t \alpha})$ is discontinuous in $t=0$ for every $\alpha\in\R^d$, and in turn $\omega$ is not regular.
\end{lemma}
\proof
Using the translation invariance of $\omega$ and the Weyl commutation relations, one infers that
$$
\begin{aligned}
\omega(u_\alpha v_\beta)\; &=\; \omega(v_\lambda u_\alpha v_\beta v_\lambda^*)\\
 &=\; \omega(\expo{-\ii\alpha\cdot\lambda}u_\alpha v_\lambda v_\lambda^*v_\beta)\\
 &=\; \expo{-\ii\alpha\cdot\lambda} \omega(u_\alpha v_\beta)
\end{aligned}
$$
for all $\lambda\in\R^d$.
Therefore $\omega(u_\alpha v_\beta)=0$ whenever $\alpha\neq 0$.  Fixing $\beta=0$ one gets that  $\omega(u_\alpha)=0$ for all $\alpha\neq 0$, and $\omega(u_0)=\omega({\bf 1})=1$ in view of the fact that $\omega$ is a state. This implies the non-regularity.
\qed

\begin{remark}\label{rk_rep_pair}
The final part of Lemma \ref{lemma_01} implies that in the GNS representation $\pi_\omega$ associated to $\omega$, the map $\alpha\mapsto\pi_\omega(u_\alpha)$ cannot be weakly nor strongly continuous at $\alpha = 0$. As a consequence, 
the unitaries $\pi_\omega(u_\alpha)$ do not admit infinitesimal generators.
\hfill $\blacktriangleleft$
\end{remark}

\medskip

We are now in the position to provide a precise characterization of $\s{S}_{\bb{W}}^\tau$. For that let us introduce the notation $\s{S}_{\bb{V}}$
for the space of states of the $C^*$-subalgebra $\bb{V}$. The next result is a refinement of  \cite[Corollary 2.15]{beaume-manuceau-pellet-sirugue-74} for the case of space translations. In particular it provides the key ingredient for the precise description of the elements of $\s{S}_{\bb{W}}^\tau$ provided in Corollary \ref{cor:trasnl_st}. Let 
\[
\s{P}_\bb{W}^\tau\;:=\;\s{S}_{\bb{W}}^\tau\cap \s{P}_\bb{W}
\]
be the set of translationally invariant pure states.

\begin{proposition}\label{prop_inv_sts}
There is an homeomorphism  $\s{S}_{\bb{W}}^\tau\simeq\s{S}_{\bb{V}}$ given as follows: every $\widetilde{\omega}\in \s{S}_{\bb{V}}$ defines a state $\omega \in \s{S}_{\bb{W}}^\tau$ through the prescription
\begin{equation}\label{eq_st_inv_01}
\omega(a)\;:=\;\widetilde{\omega}(\langle a\rangle)\;,\qquad \forall\; a\in \bb{W}\;,
\end{equation}
and all the elements of $\s{S}_{\bb{W}}^\tau$ are of this form. Moreover,  this  homeomorphism restricts to an homeomorphism $\s{P}_\bb{W}^\tau\simeq \s{P}_\bb{V}$.
\end{proposition}
\proof
 We first check that states of the form \eqref{eq_st_inv_01}
 belong to $\s{S}_{\bb{W}}^\tau$. Indeed, for $\lambda\in\R^d$,
\begin{align*}
    \omega\circ\tau_\lambda(a) &= \widetilde\omega(\langle \tau_\lambda( a )\rangle ) =
    \widetilde\omega (\langle a \rangle )  = \omega (a).
\end{align*}
Hence, such $\omega$ defines an element of $\s{S}_{\bb{W}}^\tau$.
Let us now show that any element $\omega\in \s{S}_{\bb{W}}^\tau$ is of the  form \eqref{eq_st_inv_01}. By lemma \ref{lemma_01} and using the 
representation \eqref{eq:exp_l2} and
continuity of the states, one notes that, initially for $a\in\bb{W}_0$ and then extending,
$$\omega(a)\; =\; \sum_{(\alpha,\beta)\in\Sigma_{g_a}} g_a(\alpha,\beta) \omega(u_\alpha v_\beta) \;=\; \sum_{\beta\in\Sigma_{g_a}} g_a(0,\beta) \omega(v_\beta) \;=\; \omega (\langle a \rangle)$$
where the equality 
\[
\langle a \rangle\;=\;\sum_{\beta\in\Sigma_{g_a}} g_a(0,\beta) v_\beta
\]
is a direct consequence of Proposition \ref{prop_inv_tau_charac}.  Thus, defining the restriction $\widetilde\omega := \omega|_{\bb{V}} \in \s{S}_{\bb{V}}$, we obtain the desired structure. 
The bicontinuity of this bijection  can be checked directly in the convergence of nets, since
\begin{align*}
    \widetilde{\omega}_i \to \widetilde{\omega} &\;\iff\; \widetilde{\omega}_i(a) \to \widetilde{\omega}(a) \,, &&\forall a\in \bb{V} \\ 
    &\;\iff\; \widetilde{\omega}_i(\langle a\rangle) \to \widetilde{\omega}(\langle a \rangle) \,, &&\forall a\in \bb{W} \\
    &\;\iff\; \omega_i \to \omega\;.&&
\end{align*}
For the last part let us start by assuming that $\widetilde\omega$ is not pure. Then there are two distinct elements $\widetilde\omega_1,\widetilde\omega_2\in \s{S}_{\bb{V}}$ and a $t\in (0,1)$ such that 
\[
\widetilde\omega(a)\;=\;t\widetilde\omega_1(a)+(1-t)\widetilde\omega_2(a)\;, \qquad \forall\; a\in\bb{V}\;.
\]
However, this translates to 
\[
\omega(a)\;=\;t \omega_1(a)+(1-t) \omega_2(a)\;, \qquad \forall\; a\in\bb{W}\;.
\]
where $\omega_j(a):=\widetilde\omega_j(\langle a\rangle)$ and $j=1,2$. Then $\omega$ cannot be pure. On the other hand if $\widetilde\omega$ is pure, one has that
\begin{equation}\label{eq_im_purit}
1\;=\;\widetilde\omega({\bf 1})\;=\;\widetilde\omega(v_\beta v_\beta^*)\;=\;\widetilde\omega(v_\beta)\widetilde\omega(v_\beta^*)\;=\;|\widetilde\omega(v_\beta)|^2
\end{equation} 
in view of the fact that $\bb{V}$ is abelian and $\widetilde\omega$ must be multiplicative. 
Let $\upsilon\in \s{S}_\bb{W}$ be any extension of $\widetilde{\omega}$ and $(\pi,\s{H},\varphi)$  the related GNS representation. Let $V := \{\varphi\}^\perp$ be  the subspace of $\s{H}$ orthogonal to $\varphi$, and $P_V$ the associated orthogonal projection. Then, for any $\beta \in\R^d$,
\begin{align*}
    1 \;=\; \|\pi(v_\beta) \varphi\|^2 &\;=\; |\langle \varphi , \pi(v_\beta) \varphi \rangle |^2 + \| P_V \pi(v_\beta) \, \varphi\|^2 \\
    &\;=\; |\widetilde\omega(v_\beta)|^2 + \| P_V \pi(v_\beta)  \varphi\|^2\\
    &\;=\; 1 + \| P_V \pi(v_\beta)  \varphi\|^2.
\end{align*}
So $ P_V \pi(v_\beta) \varphi  = 0$ and in turn $\pi(v_\beta) \varphi= c_\beta\varphi$ for some $c_\beta\in\C$ such that $|c_\beta|=1$. This immediately implies that $c_\beta=\widetilde{\omega}(v_\beta)$, namely
 $\pi(v_\beta) \varphi = \widetilde{\omega}(v_\beta) \varphi$. Using this, for any $\alpha,\beta,{\beta}'\in\R^d$,
\begin{align*}
    \upsilon(u_\alpha v_\beta) &= \langle \pi(v_{\beta'}) \varphi, \pi(v_{\beta'}) \pi(u_\alpha v_\beta) \varphi \rangle  \\
        &= \expo{-\ii \alpha\cdot {\beta'}} \langle \pi(v_{{\beta'}}) \varphi,  \pi(u_\alpha v_\beta) \pi(v_{{\beta'}}) \varphi \rangle \\
        &= \expo{-\ii \alpha\cdot \beta'} |\widetilde{\omega}(v_{\beta'}) |^2 \langle \varphi,  \pi(u_\alpha v_\beta)  \varphi \rangle \\
        &= \expo{-\ii \alpha\cdot \beta'} \upsilon (u_\alpha v_\beta).
\end{align*}
Therefore $\upsilon(u_\alpha v_\beta) = \delta_{\alpha,0}\widetilde{\omega}(v_\beta) = \omega(u_\alpha v_\beta)$, and thus $\omega$ given by \eqref{eq_st_inv_01} is the only possible extension of $\widetilde{\omega}$ to $\bb{W}$. Finally, since $\widetilde{\omega}$ is pure, it must have at least one pure extension to $\bb{W}$ \cite[2.10.2]{dixmier-77}. This implies  that $\omega$ must be pure. Therefore, we proved that the map \eqref{eq_st_inv_01} restricts to a bijection between $\s{P}_\bb{W}^\tau$ and $\s{P}_\bb{V}$. Since the map \eqref{eq_st_inv_01} is an homeomorphism, its bijective restriction must be one as well.

\medskip

According to the previous result, every  $\omega \in \s{S}_{\bb{W}}^\tau$ acts on the monomials as
\begin{equation}\label{eq:caract_st_01}
\omega(u_\alpha v_\beta)\;=\;\delta_{\alpha,0}\;\widetilde{\omega}(v_\beta)
\end{equation}
for a given $\widetilde{\omega}\in \s{S}_{\bb{V}}$. This formula is similar to the one proposed in \cite[Proposition 1]{acerbi-94}.
Hence the elements of  $\s{S}_{\bb{W}}^\tau$ are examples of  
\emph{flat non-regular states} according to their terminology.
The next information follows directly from the characterization \eqref{eq:caract_st_01}. 
\begin{corollary}\label{cor_pas_fr_dym}
Every  $\omega \in \s{S}_{\bb{W}}^\tau$  is invariant under the free dynamics $\Phi_t$
defined by
\eqref{eq:free_dyn}. 
\end{corollary}
\proof
Let us observe that
\begin{align*}
    \omega  \circ \Phi_t (u_\alpha v_\beta) 
    &\;=\; \expo{\ii\frac{t}{2}|\alpha|^2} \omega(u_\alpha v_{\beta- t\alpha}) 
    \;=\; \expo{\ii\frac{t}{2}|\alpha|^2} \delta_{\alpha,0} \;\widetilde{\omega}(v_{\beta- t\alpha}) \\
    &\;=\; \delta_{\alpha,0}\; \widetilde{\omega}(v_{\beta})
    \;=\; \omega(u_\alpha v_\beta)\;.
\end{align*}
Since the last equality is valid for every monomial $u_\alpha v_\beta$, this 
implies that
$\omega\circ \Phi_t=\Phi_t$ for every $t\in\R$.
\qed

\medskip

We need to better understand the structure of the elements in $\s{S}_{\bb{W}}^\tau$.
In view of Remark \ref{rk:ab_bhor_comp}, one has that the elements of $\s{S}_{\bb{V}}$ are in correspondence with the states of the $C^*$-algebra $C(\rr{b}(\n{R}^d))$. The Riesz-Markov-Kakutani representation theorem characterizes the state space of $C(\rr{b}(\n{R}^d))$ as the set $\s{M}_{+,1}(\rr{b}(\n{R}^d))$ of the positive, countably additive, regular Borel probability measures over $\rr{b}(\n{R}^d)$ (\cf \cite[Theorem 2.1]{hewitt-53}), equipped with the $\ast$-weak topology. A review of the crucial facts concerning the measure theory of $\rr{b}(\n{R}^d)$ is contained in Appendix \ref{app:mas_thr}. For a matter of notation let us write $\n{R}^d_{\rr{b}}$ for the space $\R^d$ endowed with the subspace topology of $\rr{b}(\n{R}^d)$ (see Appendix \ref{app:gen_fct}).
\begin{proposition}\label{cor:trasnl_st}
Every $\mu_\omega\in \s{M}_{+,1}(\rr{b}(\n{R}^d))$ defines a state
$\omega \in \s{S}_{\bb{W}}^\tau$ according to 
\[
\omega(u_\alpha v_\beta)\;=\;\delta_{\alpha,0}\;\int_{\rr{b}(\n{R}^d)}\dd\mu_\omega(\lambda) \;\xi_{-\beta}(\lambda)
\]
where $\xi_{-\beta}$ is the Gelfand transform of $v_{-\beta}$. Moreover the correspondence $\mu_\omega \mapsto \omega$ is an homeomorphism $\s{M}_{+,1}(\rr{b}(\n{R}^d)) \to \s{S}_{\bb{W}}^\tau$. Finally, the map $\beta\mapsto\omega(v_\beta)$ is continuous if and only if the support of $\mu_{\omega}$ is contained in $\n{R}^d_{\rr{b}}$
 and in such a case one has that 
\[
\omega(u_\alpha v_\beta)\;=\;\delta_{\alpha,0}\;\int_{\R^d}{\dd\widehat{\mu}_\omega(\lambda) \;\expo{-\ii \lambda\cdot\beta}}\;.
\]
where $\widehat{\mu}_\omega\in \s{M}_{+,1}(\n{R}^d)$ is the unique regular extension of $\mu_{\omega}$.
\end{proposition}
\proof
In view of the Riesz-Markov-Kakutani representation theorem,
there is an  {homeomorphism} between states $\widetilde{\omega}\in \s{S}_{\bb{V}}$ and measures $\mu_\omega\in \s{M}_{+,1}(\rr{b}(\n{R}^d))$
which provides
\[
\widetilde{\omega}(v_\beta)\;=\;\int_{\rr{b}(\n{R}^d)}\dd\mu_\omega(\lambda) \;\xi_{-\beta}(\lambda)\;.
\]
Therefore the first part of  the claim just follows by the characterization \eqref{eq:caract_st_01} which follows from Proposition \ref{prop_inv_sts}. The second part follows by Corollary \ref{corol:restr_measu}.
\qed

\medskip

\begin{remark}
It is worth observing that the canonical tracial state $\rr{t}$ defined by \eqref{eq:tr_st} provides an example of a translationally invariant state which is not semi-regular. Thereofre, the measure $\mu_\rr{t}\in \s{M}_{+,1}(\rr{b}(\n{R}^d))$ inducing its integral representation  is not supported in $\R^d$. Moreover, $\rr{t}$ is not pure. In fact  if it were pure, then by Proposition \ref{prop_inv_sts} it would restrict to a pure state of $\bb{V}$. However pure states of $\bb{V}$ meet condition \eqref{eq_im_purit} which is incompatible with $\rr{t}(v_\beta) =\delta_{\beta,0}$. \hfill $\blacktriangleleft$
\end{remark}

\subsection{States invariant under lattice translations}
In this section, we will investigate a larger class of states of $\bb{W}$ which are invariant only under discrete translations. This type of states has already been described in \cite[Theorem 3.13]{beaume-manuceau-pellet-sirugue-74} .
Let us start by fixing the notation.
Let $\{\rr{e}^1,\ldots,\rr{e}^d\}$ be a basis of $\R^d$ (orthogonality or normalization are not required) and consider the lattice
\begin{equation}\label{eq_latt}
\Gamma\;:=\;\{\gamma\in\R^d\;|\; \gamma:=\gamma_1\rr{e}^1+\ldots+\gamma_d\rr{e}^d\;,\;\;\gamma_1,\ldots,\gamma_d\in\Z\}\;\simeq\;\Z^d\;.
\end{equation}
The associated (semi-closed) unit cell is given by
\begin{equation}\label{eq:unit_cell}
Q_\Gamma\;:=\{y\in\R^d\;|\; y:=y_1\rr{e}^1+\ldots+y_d\rr{e}^d\;,\;\;y_1,\ldots,y_d\in[0,1)\}\;.
\end{equation}
Let $\n{T}_\Gamma:=\R^d/\Gamma$ be the quotient space. The classes of $\n{T}_\Gamma$ are in bijection with the points of $Q_\Gamma$, and in fact $\n{T}_\Gamma$ coincides with the $d$-dimensional (flat) torus obtained by the identification of the opposing faces of the closed cell $\overline{Q_{\Gamma}}$. We will use the notation $\beta=y_\beta+\gamma_\beta$ to refer at the unique decomposition of $\beta\in\R^d$ into $\gamma_\beta\in\Gamma$ and  $y_\beta\in Q_\Gamma$. We will also tacitly identify $y_\beta$ with the class $[\beta]\in \n{T}_\Gamma$ assuming that  $y_\beta$ is the natural representative of $[\beta]$.
The dual lattice of $\Gamma$, denoted with  $\Gamma'$,  is defined by
\begin{equation}\label{eq:dual_lattice}
    \Gamma'\;:=\;\{\gamma'\in\R^d\;|\; {\gamma'}\cdot\gamma\in 2\pi \Z\;,\;\;\forall\;\gamma\in\Gamma\}\;\simeq\;\Z^d\;.
\end{equation}
It coincides with the lattice generated by the dual basis $\{\rr{f}^1,\ldots,\rr{f}^d\}$
defined by $\rr{f}^i\cdot\rr{e}^j=2\pi\delta_{i,j}$. Its unit cell is $Q_{\Gamma'}$, although for reasons that will be clarified later, it is more relevant the so called \emph{Brillouin zone}
\[
\n{B}_\Gamma\;:=\;\R^d/\Gamma'\;\simeq\;\n{T}^d\;.
\]
As above we will use the notation $\alpha=\kappa_\alpha+\gamma'_\alpha$ for the unique decomposition of $\alpha\in\R^d$ into $\gamma'_\alpha\in\Gamma'$ and  $\kappa_\alpha\in Q_{\Gamma'}$. Also in this case the identification $\kappa_\alpha\equiv[\alpha]\in\n{B}_\Gamma$
will be tacitly used to  imply that $\kappa_\alpha$ is the natural representative of $[\alpha]$.
The points $\kappa\in \n{B}_\Gamma$ are called \emph{quasi-momenta}.

\begin{definition}[Lattice-invariant states]
A state $\omega\in\s{S}_\bb{W}$  is $\Gamma$-\textit{invariant} if 
$\omega\circ \tau_\gamma   = \omega$ for all $\gamma\in\Gamma$. The set of 
$\Gamma$-\textit{invariant}  states will be denoted with $\s{S}_{\bb{W}}^\Gamma$.
\end{definition}

\medskip

Let us start with a preliminary result similar to Lemma \ref{lemma_01}.

\begin{lemma}\label{lemma_01_01}
Let $\omega\in\s{S}_{\bb{W}}^\Gamma$ be a $\Gamma$-invariant state. Then
\begin{equation}
    \label{eccond2-2-2}
    \begin{aligned}
    \omega(u_\alpha v_\beta)\;&=\;0,\;&&\forall\;\alpha\in\R^d\setminus\Gamma'\;,\;\;\forall\;\beta\in\R^d\;. 
    \end{aligned}
   \end{equation}
As a consequence, the map $t\mapsto \omega(u_{t \alpha})$ is discontinuous in $t=0$ for every $\alpha\in\R^d$, and in turn $\omega$ is not regular. 
\end{lemma}

\proof
The same computation as in the proof of Lemma \eqref{lemma_01} provides
\[
\omega(u_\alpha v_\beta)\; =\;\expo{-\ii\alpha\cdot\gamma} \omega(u_\alpha v_\beta)\;,\qquad \forall\; \gamma\in\Gamma\;
\]
and this justifies \eqref{eccond2-2-2}. 
 Fixing $\beta=0$ one gets that  $\omega(u_\alpha)=0$ for all $\alpha\notin \Gamma'$, and $\omega(u_0)=\omega({\bf 1})=1$ in view of the fact that $\omega$ is a state.
\qed

\medskip

To have a more precise description of $\s{S}_{\bb{W}}^\Gamma$ let us repeat the strategy of Section \ref{sec_tr_inv_st}.
Let 
\[
{\rm Inv}_\tau^\Gamma(\bb{W})\;:=\;\left\{a\in\bb{W}\;|\;  \tau_\gamma(a)=a\;,\;\;\forall\; \gamma\in\Gamma\right\}
\]
be the space of elements left invariant by translations of the lattice $\Gamma$, and consider the  $C^*$-subalgebra $\bb{V}_\Gamma\subset \bb{W}$ generated solely by monomials $u_{\gamma'} v_\beta$ with $\gamma'\in\Gamma'$ and $\beta\in\R^d$.

\begin{remark}\label{rk:ab_bhor_comp_00}
In the following, we will make use of
the characterization of the    $C^*$-algebra $\bb{V}_\Gamma$ as a \emph{crossed product}. First of all, let us notice that $\bb{V}\subset\bb{V}_\Gamma$. Moreover, 
from $u_{\gamma'} v_\beta u_{\gamma'}^*=\expo{\ii \gamma'\cdot \beta}v_\beta$ one argues that the unitaries $u_{\gamma'}$  implement automorphisms of $\bb{V}$. By using the identification $\bb{V}\simeq C(\rr{b}(\n{R}^d))$ described in Remark \ref{rk:ab_bhor_comp} and given by $v_\beta\mapsto \xi_\beta$,
 one observes that $u_{\gamma'} v_\beta u_{\gamma'}^*$ is mapped to the function $\xi_\beta(\cdot+\gamma')$.
Therefore the $u_{\gamma'}$'s implement translations of the elements of $C(\rr{b}(\n{R}^d))$. This justifies the identification 
\[
\bb{V}_\Gamma\;\simeq C(\rr{b}(\n{R}^d))\rtimes\Gamma'\;.
\]
Namely, $\bb{V}_\Gamma$ coincides with the crossed product $C^*$-algebra of $C(\rr{b}(\n{R}^d))$ with respect to the action of $\Gamma'$. For more details on the concept of crossed product by a discrete group we will refer to \cite[Section 2.3]{williams-07} or to \cite[Chapter VIII]{davidson-96}. Of particular relevance is \cite[Corollary VIII.3.6]{davidson-96}. \hfill $\blacktriangleleft$
\end{remark}

\medskip

To provide a description of ${\rm Inv}_\tau^\Gamma(\bb{W})$, we will adapt the strategy used in Section \ref{sec_tr_inv_st} by introducing the \emph{ergodic $\Gamma$-mean} of an element $a\in \bb{W}$, given by
\begin{equation}\label{eq:er_mean_gamma}
\langle a\rangle_\Gamma\;:=\;\lim_{N\to+\infty}\frac{1}{|\Gamma_N|}\sum_{\gamma\in \Gamma_N}\; \tau_\gamma(a)
\end{equation} 
where $N\in\N$,  $\Gamma_N:=\Gamma\cap\Lambda_N$ and $|\Gamma_N|=(2N+1)^d$ is the cardinality of $\Gamma_N$. This formula is similar to \eqref{eq:er_mean} by replacing the Lebesgue measure on $\R^d$ with the counting measure on $\Gamma$.
In this case one again has that $\langle a\rangle_\Gamma\in{\rm Inv}_\tau^\Gamma(\bb{W})$ and the linear map $\langle \cdot\rangle_\Gamma:\bb{W}\to {\rm Inv}_\tau^\Gamma(\bb{W})$ results to be norm continuous in view of the inequality $\|\langle a\rangle_\Gamma\|\leqslant \|a\|$.

\begin{proposition}\label{prop_gamma_inv_sts}
It holds true that ${\rm Inv}_\tau^\Gamma(\bb{W})= \bb{V}_\Gamma$.
\end{proposition}
\proof
The argument follows identically the proof of Proposition \ref{prop_inv_tau_charac}. The main point is the proof of the equality  
 
\[
\langle u_\alpha v_\beta\rangle_\Gamma
\;=\;
\left(\lim_{N\to+\infty}\frac{1}{|\Gamma_N|}\sum_{\gamma\in \Gamma_N} \expo{-\ii\alpha\cdot\gamma}\right) u_\alpha v_\beta
\;=\;
\chi_{\Gamma'}(\alpha) u_\alpha v_\beta\;\in\; \bb{V}_\Gamma\; 
\]
where $\chi_{\Gamma'}$ denotes the characteristic function of the dual lattice $\Gamma'$.
\qed

\medskip

Let $\s{S}_{\bb{V}_\Gamma}$ be the space of states of the $C^*$-subalgebra $\bb{V}_\Gamma$. The next result is similar to Proposition \ref{prop_inv_sts}
  for the case of $\Gamma$-invariant states. For that
let us introduce the set 
\[
\s{P}_\bb{W}^\Gamma\;:=\;\s{S}_{\bb{W}}^\Gamma\cap \s{P}_\bb{W}
\]
of $\Gamma$-invariant pure states.
\begin{proposition}\label{prop_inv_sts_oo}
There is an homeomorphism  $\s{S}_{\bb{W}}^\Gamma\simeq\s{S}_{\bb{V}_\Gamma}$ given as follows:
  every $\widetilde{\omega}\in \s{S}_{\bb{V}_\Gamma}$ defines a state $\omega \in \s{S}_{\bb{W}}^\Gamma$ through the prescription
\begin{equation}\label{eq:FFRRTT}
\omega(a)\;:=\;\widetilde{\omega}(\langle a\rangle_\Gamma)\;,\qquad \forall\; a\in \bb{W}\;,
\end{equation}
and all the elements of $\s{S}_{\bb{W}}^\Gamma$ are of this form. Moreover,  this  homeomorphism restricts to a topological embedding
$\s{P}_\bb{W}^\Gamma \hookrightarrow\s{P}_{\bb{V}_\Gamma}$.
\end{proposition}
\proof
The argument is exactly the same as the proof of Proposition \ref{prop_inv_sts}. The difference is that $\widetilde\omega\in \s{P}_{\bb{V}_\Gamma}$ doesn't imply in general that $\omega\in\s{P}_\bb{W}^\Gamma$ with $\omega$ given by \eqref{eq:FFRRTT}.
\qed

\medskip

By combining the previous result with the computation in the proof of Proposition \ref{prop_gamma_inv_sts}, one gets that every $\omega \in \s{S}_{\bb{W}}^\Gamma$ acts on the monomials as
\begin{equation}\label{eq:caract_st_01_gamma}
\omega(u_\alpha v_\beta)\;=\;\chi_{\Gamma'}(\alpha)\;\widetilde{\omega}(u_\alpha v_\beta)
\end{equation}
for a given $\widetilde{\omega}\in \s{S}_{\bb{V}_\Gamma}$.
Again, this formula is   similar to the one proposed in \cite[Proposition 1]{acerbi-94}.

\medskip

In order to have an equivalent of Corollary \ref{cor:trasnl_st} for a structural description of $\s{S}_{\bb{W}}^\Gamma$ one should be able to describe the state space $\s{S}_{\bb{V}_\Gamma}$. In view of Remark \ref{rk:ab_bhor_comp_00}, the latter task is equivalent to the description of the state space of the crossed product $C(\rr{b}(\n{R}^d))\rtimes\Gamma'$. While this is an interesting question, for the purposes of this work we will only focus on the description of the set
\[
 \s{P}_\bb{W}^{\Gamma,\beta}\;:=\;\{ \omega\in\s{P}_{\bb{W}}^\Gamma\;|\; \omega\;\;\text{is semi-regular in the parameter}\;\;\beta\}
\]
of $\Gamma$-invariant pure states such that the mappings $\beta\mapsto \omega(u_\alpha v_\beta)$ are continuous. The elements of  $ \s{P}_\bb{W}^{\Gamma,\beta}$ will be called \emph{Bloch-wave states}.
A description of these states is already given  in \cite[Section 3]{acerbi-94} and in  \cite[Theorem 3.13]{beaume-manuceau-pellet-sirugue-74}. Let us start with a preliminary result.
\begin{proposition}\label{prop_intras}
Let $\omega\in \s{P}_\bb{W}^{\Gamma}$ and $(\pi, \mathfrak{h}, \psi)$ its
 associated GNS representation. Then
 \begin{equation}\label{eq:phasDD}
\pi(v_\gamma)\psi\; =\;  \expo{-\ii\kappa\cdot\gamma} \psi\;\qquad \forall \gamma\in\Gamma
\end{equation}
for some fixed $\kappa\in\n{B}_\Gamma$, and one refers to $\kappa$ as the \emph{quasi-momentum} of $\omega$.
\end{proposition}
\proof
Let $\rr{h}_0:=\pi(\bb{W})\psi$ be the dense subspace spanned by the cyclic vector inside $\rr{h}$. 
For $\gamma\in\Gamma$ let us define the operator $T(\gamma)$ initially on $\rr{h}_0$ by 
\[
T(\gamma) \pi(a) \psi\;: =\; \pi(\tau_\gamma(a)) \psi\;, \qquad  \quad \forall a \in \bb{W}\;.
\]
By using the $\Gamma$-invariance of $\omega$ one infers that $\|T(\gamma) \pi(a) \psi\|^2=\| \pi(a) \psi\|^2$ for every $a \in \bb{W}$. Thus, the density of $\rr{h}_0$
implies that $T(\gamma)$ extends to a bounded operator on $\rr{h}$ of norm one.
From its very definition it follows that $T(-\gamma)=T(\gamma)^{-1}$. Finally the equality
$\omega(\tau_\gamma(b^*)a)=\omega(b^*\tau_{-\gamma}(a))$ is valid for every $a,b\in \bb{W}$
 in view of the fact the $\Gamma$-invariance of $\omega$ implies that  $T(-\gamma)=T(\gamma)^*$. As a result one obtains that $T(\gamma)$ is a unitary operator.
Of course, $T(\gamma) \psi = \psi$.  Furthermore,
\[
T(\gamma) \pi(a) T(\gamma)^* \pi(b) \psi\; =\;T(\gamma)\pi(a\tau_{-\gamma}(b))\psi
\;=\;\pi(\tau_{\gamma}(a)b)\psi\;=\;\pi(\tau_{\gamma}(a))\pi(b)\psi
\]
for every $b\in\bb{W}$. By density this implies that 
$T(\gamma)\pi(a)T(\gamma)^* = \pi(\tau_\gamma(a))$ for every $a\in\bb{W}$.
 Consequently,
\[
T(\gamma)^* \pi (v_\gamma) \pi(a)\; =\; T(\gamma)^* \pi(\tau_\gamma (a)) \pi (v_\gamma)\; =\; \pi(a) T(\gamma)^* \pi (v_\gamma)\;,
\]
so that $T(\gamma)^* \pi  (v_\gamma)$ commutes with every element of $\pi (\bb{W})$. Since the representation is irreducible (purity of $\omega$) and faithful (simplicity of $\bb{W}$) it follows that $T(\gamma)^*\pi (v_\gamma)=c_\gamma {\bf 1}$  for a  $c_\gamma\in\C$ such that $|c_\gamma|=1$. Let us observe that
\[
c_\gamma c_{\xi} {\bf 1}\;=\;T(\gamma)^* \pi  (v_\gamma)T(\xi)^* \pi  (v_{\xi})
\;=\;T(\gamma+\xi)^* \pi  (v_{\gamma+\xi})\;=\;c_{\gamma+\xi} {\bf 1}
\]
for every $\gamma,\xi\in\Gamma$. This implies that 
the map $\gamma\mapsto c_\gamma$ is a character of $\Gamma$, namely
$c_\gamma =\expo{-\ii \kappa \cdot \gamma}$ for some fixed $\kappa\in\mathbb{B}_\Gamma$. This implies
\begin{equation*}
\pi(v_\gamma)\psi\; =\; T(\gamma) T(\gamma)^* \pi(v_\gamma)\psi\; =\; \expo{-\ii \kappa \cdot \gamma}T(\gamma)\psi\; =\; \expo{-\ii\kappa\cdot\gamma} \psi\;.
\end{equation*}
as claimed.
\qed

\medskip

The last result shows that the elements of  $\s{P}_\bb{W}^{\Gamma}$ are differentiated by the points $\kappa\in\n{B}_\Gamma$. 
However, there is no reason why a given quasi-momentum $\kappa$ should specify a unique element of
  $\s{P}_\bb{W}^{\Gamma}$. In particular the GNS representations of states corresponding to distinct values of the quasi-momentum cannot be related by a unitary transformation in view of \eqref{eq:phasDD}.
{In the following we will refine this result into a characterization of the GNS representations of the semi-regular pure states. This will allow us to fully describe the set  $\s{P}_\bb{W}^{\Gamma,\beta}$}.
For that we need first to introduce some notation.   
Consider the Hilbert space $\rr{h}_\Gamma:=L^2(\n{T}_\Gamma,\dd\nu)$ where $\n{T}_\Gamma:=\R^d/\Gamma$ and 
 $\dd\nu:=|Q_\Gamma|^{-1}\dd y$ (with $\dd y$  the Lebesgue measure of $\R^d$) its normalized Haar measure. Observe that 
the Pontryagin dual of $\n{T}_\Gamma$ coincides with $\Gamma'$. Consider two families of operators on  $\rr{h}_\Gamma$ defined on  $f\in\rr{h}_\Gamma$ by
\begin{equation}\label{eq_SS-YY}
\begin{aligned}
(S_\beta f)(y) &\;:=\; f(y-\beta) \\
(F_{\gamma'} f)(y) &\;:=\; \expo{\ii \gamma'\cdot y}f(y)
\end{aligned}
\end{equation}
for every $\gamma'\in\Gamma'$ and $\beta\in\R^d$. 
Let $\beta=y_\beta+\gamma_\beta$ the unique decomposition of $\beta$ in $\gamma_\beta\in\Gamma$ and $y_\beta\in Q_\Gamma$, and
$y_\beta\equiv[\beta]\in \n{T}_\Gamma$ the associated class. 
Then $S_\beta=S_{[\beta]}$ for all $\beta\in\R^d$, and the maps $[\beta]\mapsto S_{[\beta]}$ and $\gamma'\mapsto F_{\gamma'}$ provide strongly continuous unitary representations of the groups $\n{T}_\Gamma$ and $\Gamma'$, respectively. 
Moreover, $F_{\gamma'}S_\beta=\expo{\ii\gamma'\cdot\beta} S_\beta F_{\gamma'}$ for every $\gamma'\in\Gamma'$ and $\beta\in\R^d$ and such commutation relations are compatible with the restriction $\beta\mapsto[\beta]$. 
With this in mind, for every $\kappa\in\n{B}_\Gamma$, we may define a representation $\rho_\kappa:\bb{V}_\Gamma\to \bb{B}(\rr{h}_\Gamma)$ by
\begin{equation*}
\begin{aligned}
{\rho}_\kappa (v_\beta)  &\;:=\; \expo{-\ii\kappa\cdot \beta} S_\beta \;,\qquad
{\rho}_\kappa(u_{\gamma'})  \;:=\; F_{\gamma'}\,.
\end{aligned}
\end{equation*}
Evidently, if $\kappa\neq\kappa'$ then the representations ${\rho}_\kappa$ and ${\rho}_{\kappa'}$ are not unitarily equivalent.

\begin{lemma}\label{lemma:GNS_quasimomentum_kappa}
    Let $\omega\in \s{P}_\bb{W}^{\Gamma,\beta}$, with quasi-momentum $\kappa\in\n{B}_\Gamma$. Denote its restriction to $\bb{V}_\Gamma$ by $\widetilde{\omega}\in \s{P}_{\bb{V}_\Gamma}$, and let $({\pi}_{\widetilde{\omega}}, {\mathfrak{h}}_{\widetilde{\omega}}, \psi_{\widetilde{\omega}})$ be the GNS representation of $\widetilde{\omega}$.
    Then, there exists a unitary operator $\s{U}: {\rr{h}}_{\widetilde{\omega}}\to \rr{h}_\Gamma$ such that
\begin{equation*}
\begin{aligned}
\s{U} {\pi}_{\widetilde{\omega}}(a)  \s{U}^{-1} \;&=\; \rho_\kappa(a) \,, \qquad \forall a\in\bb{V}_\Gamma.
\end{aligned}
\end{equation*}
Thus, the representation $\rho_\kappa:\bb{V}_\Gamma\to\bb{B}(\rr{h}_\Gamma)$, with cyclic vector $f_{\widetilde{\omega}}:=\s{U}\psi_{\widetilde{\omega}}$, is unitarily equivalent to the GNS representation of $\widetilde{\omega}$.
\end{lemma}
\proof
Let $\theta_\alpha$ be the automorphism of $\bb{W}$ given by 
$\theta_\alpha(a) := u_\alpha a u_\alpha^*$ for every $a\in\bb{W}$ and $\alpha\in\R^d$ (momentum translations). It follows that 
$\theta_\alpha:\bb{V}_\Gamma\to\bb{V}_\Gamma$ restricts to an automorphism of $\bb{V}_\Gamma$ for every $\alpha\in\R^d$.
Therefore, the map $\sigma: \bb{V}_\Gamma\to\bb{B}({\mathfrak{h}}_{\widetilde{\omega}})$ defined by $\sigma:={\pi}_{\widetilde{\omega}}\circ\theta_\kappa$ 
is still an irreducible  representation, and  
$\upsilon := \omega \circ \theta_\kappa$ and $\widetilde{\upsilon} := \upsilon |_{\bb{V}_\Gamma} = \widetilde{\omega}\circ \theta_\kappa$ are still pure states. Note that
$\sigma$ provides the GNS representation of $\widetilde{\upsilon}$ with respect the same cyclic vector $\psi_{\widetilde{\omega}}$.
For every $a,b\in \bb{V}_\Gamma$ one gets
\[
\langle\sigma(a)\psi_{\widetilde{\omega}}, \sigma(v_\gamma)\sigma(b)\psi_{\widetilde{\omega}}\rangle_{{\mathfrak{h}}_{\widetilde{\omega}}}\;=\;\widetilde{\upsilon}(a^*v_\gamma b)\;=\; \upsilon(a^*bv_\gamma) 
\]
where in the last equation we used $v_\gamma b=bv_\gamma$ and the fact that $\upsilon$ extends $\widetilde{\upsilon}$. Since $\theta_\kappa(v_\gamma)=\expo{\ii \kappa\cdot \gamma} v_\gamma$,
\[
\langle\sigma(a)\psi_{\widetilde{\omega}}, \sigma(v_\gamma)\sigma(b)\psi_{\widetilde{\omega}}\rangle_{{\mathfrak{h}}_{\widetilde{\omega}}}\;=\;\expo{\ii \kappa\cdot \gamma}\omega(\theta_\kappa(a^*b)v_\gamma)\;.
\]
As a consequence, one infers that 
\[
\begin{aligned}
\langle\sigma(a)\psi_{\widetilde{\omega}}, \sigma(v_\gamma)\sigma(b)\psi_{\widetilde{\omega}}\rangle_{{\mathfrak{h}}_{\widetilde{\omega}}}\;&=\;\omega(\theta_\kappa(a^*b))\;=\;\upsilon(a^*b)\\
&=\;\widetilde{\upsilon}(a^*b)\;=\;\langle\sigma(a)\psi_{\widetilde{\omega}}, \sigma(b)\psi_{\widetilde{\omega}}\rangle_{{\mathfrak{h}}_{\widetilde{\omega}}}\;
\end{aligned}
\]
where the first equality just follows from an application   of Proposition \ref{prop_intras}. By density it follows that
 $\sigma(v_\gamma)={\bf 1}$, or equivalently $\sigma(v_{\beta+\gamma})=\sigma(v_{\beta})$, for every $\gamma\in\Gamma$. From this and the $\beta$-regularity of $\omega$, one concludes that $\sigma(v_{\beta}) = \sigma(v_{[\beta]})$ is a continuous representation of the group $\n{T}_\Gamma$. Since the hypothesis of the Stone-von Neumann-Mackey Theorem \cite[Theorem 1]{mackey-49} are satisfied there is a unitary operator $\s{U}: {\mathfrak{h}}_{\widetilde{\omega}}\to \rr{h}_\Gamma$ such that
\begin{equation*}
\begin{aligned}
\s{U} \sigma (v_y)  \s{U}^{-1} &\;=\; S_y \;,\qquad
\s{U} \sigma (u_{\gamma'})  \s{U}^{-1}  \;=\; F_{\gamma'}.
\end{aligned}
\end{equation*}
Finally, since $\theta_\kappa(u_{\gamma'} v_\beta)=\expo{\ii \kappa\cdot\beta} u_{\gamma'} v_\beta$, one gets $\sigma(u_{\gamma'} v_\beta)=\expo{\ii \kappa\cdot\beta} \pi_{\widetilde{\omega}}(u_{\gamma'} v_\beta)$. Therefore, one concludes that $\s{U} \pi_{\widetilde{\omega}} (u_{\gamma'} v_\beta) \s{U}^{-1} = \expo{-\ii \kappa\cdot\beta} F_{\gamma'} S_\beta = \rho_\kappa (u_{\gamma'} v_\beta)$ for every $\gamma'\in\Gamma'$ and $\beta\in\R^d$, and this
 implies the desired result.
\qed

\medskip

We now have the tools needed to describe the GNS representation of a state
$\omega\in \s{P}_\bb{W}^{\Gamma,\beta}$.
For the next result, which compares to \cite[Theorem 2.16]{beaume-manuceau-pellet-sirugue-74}, we need the (non-separable) Hilbert space $\ell^2(\n{B}_\Gamma)$ of functions $t:\n{B}_\Gamma\to\C$ with countable support and $\ell^2$-summability. A canonical basis is provided by the Dirac deltas $\delta_{[\nu]}$ with $[\nu]\in\n{B}_\Gamma$. 
In the following we will use $\alpha=\kappa_\alpha+\gamma'_\alpha$ for the unique decomposition of $\alpha\in\R^d$ in $\gamma'_\alpha\in\Gamma'$ and $\kappa_\alpha\in Q_{\Gamma'}$ and we will tacitly identify $\kappa_\alpha$ with its class $[\alpha] \in \n{B}_\Gamma$. Let us introduce the Hilbert space $\rr{H}_\Gamma := \ell^2(\n{B}_\Gamma)\otimes \rr{h}_\Gamma$ and  for every $f\in \rr{h}_\Gamma$ consider the family of vectors $\{\Psi^f_{\nu,\sigma}\}_{\nu,\sigma\in\R^d}\subset \rr{H}_\Gamma$ defined by
\[
\Psi^f_{\nu,\sigma}\;:=\; \delta_{[\nu]} \otimes F_{\gamma'_\nu} S_\sigma f\;.
\]
Observing that $F_{\gamma'_\nu} S_\sigma=\rho_0(u_{\gamma'_\nu}v_\sigma)$, and that the representation $\rho_0$ is irreducible (Lemma \ref{lemma:GNS_quasimomentum_kappa}), it follows from \cite[Proposition 2.3.8]{bratteli-robinson-87} that any $f$ is cyclic for $\rho_0$, and in turn that for any fixed $f$, the family $\Psi^f_{\nu,\sigma}$ spans a dense set in  $\rr{H}_\Gamma$.

\begin{proposition}\label{prop:GNS_bloch_wave_sts}
Consider the representation $\rho:\bb{W}\to\bb{B}(\rr{H}_\Gamma)$ initially defined by
\begin{equation*}
\begin{aligned}
     \rho(u_\alpha)\Psi^f_{\nu,\sigma} \;&:=\; \Psi^f_{\nu+\alpha,\sigma}\;,\\
     \rho(v_\beta) \Psi^f_{\nu,\sigma}\;&:=\; \expo{-\ii\nu\cdot\beta}\Psi^f_{\nu,\sigma+\beta}\;.
\end{aligned}
\end{equation*}
on the family $\{\Psi^f_{\nu,\sigma}\}_{\nu,\sigma\in\R^d}$,  and then extended by linearity and density to the  entire space $\rr{H}_\Gamma$. The definition of $\rho$ does not depend on the choice of $f\in\rr{h}_\Gamma$.
Let $\omega\in \s{P}_\bb{W}^{\Gamma,\beta}$ with quasi-momentum $\kappa$ (identified by its representative in the unit cell),  $\widetilde{\omega}:=\omega|_{\bb{V}_\Gamma}$ its restriction and $({\rho}_\kappa, \mathfrak{h}_\Gamma, f_{\widetilde{\omega}})$  the  GNS representation of 
$\widetilde{\omega}$ described in Lemma \ref{lemma:GNS_quasimomentum_kappa}. Then $(\rho,\rr{H}_\Gamma,\Psi_{\kappa,0}^{f_{\widetilde{\omega}}})$ is (up to unitary equivalence) the GNS representation of $\omega$.
\end{proposition}
\proof 
The fact that $\rho$ is a representation follows just from checking that $ \rho(u_\alpha)$ and $ \rho(v_\beta)$ meet the Weyl relations. The fact that the definition is independent of the specific vector $f\in\rr{h}_\Gamma$ can be readily verified on the dense subspace $\rho_0(\bb{V}_\Gamma)f$ through a cumbersome but straightforward calculation, and then extended to all of $\rr{h}_\Gamma$.
The key observation is that
\[
 \Psi^{\rho_0(u_{\gamma'}v_\eta)f}_{\nu,\sigma} \;=\;\expo{-\ii \gamma'\cdot\sigma} \Psi^f_{\nu+\gamma',\sigma+\eta}
\]
for every $\gamma'\in\Gamma'$ and $\eta\in\R^d$. Therefore
\[
\begin{aligned}
\rho(u_\alpha v_\beta)  \Psi^{\rho_0(u_{\gamma'}v_\eta)f}_{\nu,\sigma} &=\;\expo{-\ii \gamma'\cdot\sigma}\left(\rho(u_\alpha v_\beta)\Psi^f_{\nu+\gamma',\sigma+\eta}\right)\\
&=\;\expo{-\ii \gamma'\cdot\sigma}\left(\expo{-\ii (\nu+\gamma')\cdot\beta}\Psi^f_{\nu+\gamma'+\alpha,\sigma+\eta+\beta}\right)\\
&=\;\expo{-\ii \nu\cdot\beta}\left(\expo{-\ii \gamma'\cdot(\sigma+\beta)}\Psi^f_{(\nu+\alpha)+\gamma',(\sigma+\beta)+\eta}\right)\\&=\;
\expo{-\ii \nu\cdot\beta}\Psi^{\rho_0(u_{\gamma'}v_\eta)f}_{\nu+\alpha,\sigma+\beta}
\end{aligned}
\]
Therefore, the definition of $\rho$ is the same when replacing $f$ by 
$\rho_0(u_{\gamma'}v_\eta)f$. Since $\Psi^{f+g}_{\nu,\sigma}=\Psi^{f}_{\nu,\sigma}+\Psi^{g}_{\nu,\sigma}$ for every $f,g\in\rr{h}_\Gamma$, one obtains by linearity that the definition of $\rho$ is the same when replacing $f$ by 
$\rho_0(a)f$ with $a\in\bb{V}_\Gamma$ a finite linear combination of monomials $u_{\gamma'}v_\eta$. Now fix any $g\in \rr{h}_\Gamma$, $\epsilon>0$, and let $a\in\bb{V}_\Gamma$ a finite linear combination of monomial such that 
 $\|\rho_0(a)f - g\|_{\rr{h}_\Gamma}<\epsilon/2$.
A straightforward computation shows that
\begin{align*}
    \|\rho(u_\alpha v_\beta)\Psi^g_{\nu,\sigma} - \expo{-\ii \nu \cdot\beta} \Psi^{g}_{\nu+\alpha,\sigma+\beta}\|_{\rr{H}_\Gamma} \;&\leq\;  \|\rho(u_\alpha v_\beta)\left(\Psi^g_{\nu,\sigma} - \Psi^{\rho_0(a)f}_{\nu,\sigma}\right)\|_{\rr{H}_\Gamma} \\
    &+ \|\expo{-\ii \nu \cdot\beta}\big( \Psi^{\rho_0(a)f}_{\nu+\alpha,\sigma+\beta} -  \Psi^{g}_{\nu+\alpha,\sigma+\beta}\big)\|_{\rr{H}_\Gamma} \\
    &=\; 2 \| \Psi^g_{0,0} -  \Psi^{\rho_0(a)f}_{0,0}\|_{\rr{H}_\Gamma}\\
    &=\; 2 \| g - \rho_0(a)f\|_{\rr{h}_\Gamma} \; <\; \epsilon \,.
\end{align*}
This shows  that
\[
\rho(u_\alpha v_\beta)\Psi^g_{\nu,\sigma} \;=\; \expo{-\ii \nu \cdot\beta} \Psi^{g}_{\nu+\alpha,\sigma+\beta}\;,
\]
which means that  the definition of $\rho$ is  the same when replacing $f$ by any other $g\in\rr{h}_\Gamma$.
Since $\omega\in \s{P}_\bb{W}^{\Gamma,\beta}$, then also $\upsilon:=\omega\circ\theta_\kappa$ is still in $\s{P}_\bb{W}^{\Gamma,\beta}$ and 
\[
\upsilon(v_\gamma)\; =\; \omega(\theta_\kappa(v_\gamma))\; =\; \expo{\ii \kappa \cdot\gamma} \omega(v_\gamma) \;=\; 1\;
\]
in view of \eqref{eq:phasDD}. Therefore, $\upsilon$ is a state with quasi-momentum $0$.
By Lemma \ref{lemma:GNS_quasimomentum_kappa}, $({\rho}_0, \mathfrak{h}_\Gamma, f_{\widetilde{\omega}})$ is a GNS triplet for $\widetilde{\upsilon} := \upsilon|_{\bb{V}_\Gamma}$. Let $\Psi_{0,0}^{f}=\delta_{0} \otimes f$ with $f\in\rr{h}_\Gamma$
and observe that
\[
 \rho(u_\alpha v_\beta) \Psi_{0,0}^{f}\;=\;\delta_{[\alpha]} \otimes \rho_0(u_{\gamma'_\alpha}v_\beta)f\;.
\]
Therefore $\Psi_{0,0}^{f}$ is cyclic for $\rho(\bb{W})$ for every $f\in\rr{h}_\Gamma$.
In particular
\[
\begin{aligned}
\left\langle \Psi_{0,0}^{f_{\widetilde{\omega}}} , \rho(u_\alpha v_\beta)\Psi_{0,0}^{f_{\widetilde{\omega}}} \right\rangle_{\rr{H}_\Gamma}\;&=\;\langle\delta_{0},\delta_{[\alpha]}\rangle_{\ell^2(\n{B}_\Gamma)}\langle f_{\widetilde{\omega}},\rho_0(u_{\gamma'_\alpha}v_\beta)f_{\widetilde{\omega}}\rangle_{\rr{h}_\Gamma}\\
&=\;\chi_{\Gamma'}(\alpha)\;\widetilde{\upsilon} (u_{\gamma'_\alpha}v_\beta)\\
&=\;\chi_{\Gamma'}(\alpha)\;\widetilde{\upsilon} (u_{\alpha}v_\beta)\;=\;\upsilon (u_{\alpha}v_\beta).
\end{aligned}
\]
Hence  $(\rho,\rr{H}_\Gamma,\Psi_{0,0}^{f_{\widetilde{\omega}}})$ is a GNS triplet for $\upsilon$. Evidently $\Psi_{\kappa,0}^{f_{\widetilde{\omega}}}=\rho(u_\kappa)\Psi_{\kappa,0}^{f_{\widetilde{\omega}}}=\delta_{\kappa} \otimes f_{\widetilde{\omega}}$ (with $\kappa\in Q_{\Gamma'}$   in the unit cell)
is also cyclic for $\rho$. Finally,
\begin{align*}
    \omega(u_\alpha v_\beta)  &=\; \upsilon(u_\kappa^* u_\alpha v_\beta u_\kappa) \\
    &=\; \left\langle \Psi_{0,0}^{f_{\widetilde{\omega}}} , \rho(u_\kappa^* u_\alpha v_\beta u_\kappa)\; \Psi_{0,0}^{f_{\widetilde{\omega}}} \right\rangle_{\rr{H}_\Gamma} \\
    &=\; \left\langle \rho(u_\kappa) \Psi_{0,0}^{f_{\widetilde{\omega}}} , \rho(u_\alpha v_\beta)\;  \rho(u_\kappa) \Psi_{0,0}^{f_{\widetilde{\omega}}}\right\rangle_{\rr{H}_\Gamma} \\
    &=\; \left\langle \Psi_{\kappa,0}^{f_{\widetilde{\omega}}}  , \rho(u_\alpha v_\beta)\; \Psi_{\kappa,0}^{f_{\widetilde{\omega}}}\right\rangle_{\rr{H}_\Gamma},
\end{align*}
and this  proves the result.
\qed

\medskip

We are finally in position to describe the elements of $\s{P}_\bb{W}^{\Gamma,\beta}$. Let $\bb{P}(\rr{h}_\Gamma)$ be the set of orthogonal projections of $\rr{h}_\Gamma$,  ${\rm Tr}_{\rr{h}_\Gamma}$ the canonical trace of $\rr{h}_\Gamma$, and
\[
\bb{G}_1(\rr{h}_\Gamma)\;:=\;\left\{P\in \bb{P}(\rr{h}_\Gamma)\;|\; {\rm Tr}_{\rr{h}_\Gamma}(P)=1\right\}
\]
the set of 1-dimensional orthogonal projections of $\rr{h}_\Gamma$. This space is known as the rank-1 Hilbert Grassmannian of $\rr{h}_\Gamma$, and provides a model for the projective (or ray) space of $\rr{h}_\Gamma$. 

\begin{proposition}[Bloch-wave states]\label{cor_pas_bet-Gamma}
Every element of $\s{P}_\bb{W}^{\Gamma,\beta}$ is of the form
\begin{equation}\label{eq_st_pur_gamm_bl}
\omega_{(\kappa,P)}(u_\alpha v_\beta) \;:=\; \chi_{\Gamma'}(\alpha)\; \expo{-\ii \kappa\cdot \beta }{\rm Tr}_{\rr{h}_\Gamma}(PF_{\alpha}S_\beta)\;
\end{equation}
where $(\kappa,P)\in Q_{\Gamma'} \times \bb{G}_1(\rr{h}_\Gamma)$. This correspondence is bijective.
\end{proposition}
\proof
Let $\omega\in \s{P}_\bb{W}^{\Gamma,\beta}$ with quasi-momentum $[\kappa]\in\n{B}_\Gamma$ identified by its value $\kappa\in Q_{\Gamma'}$ in the unit cell. By Proposition \ref{prop:GNS_bloch_wave_sts}, there is a $f_{\widetilde{\omega}}\in\rr{h}_\Gamma$ such that
\begin{align*}
    \omega(u_\alpha v_\beta) \;&=\; 
    \left\langle \Psi_{\kappa,0}^{ f_{\widetilde{\omega}}} , \rho(u_\alpha v_\beta) \Psi_{\kappa,0}^{ f_{\widetilde{\omega}}}\right\rangle_{\rr{H}_\Gamma}  \\
    &=\; \left\langle \Psi_{\kappa,0}^{ f_{\widetilde{\omega}}} ,\, \expo{-\ii \kappa\cdot \beta} \Psi_{\kappa+\alpha,\beta}^{ f_{\widetilde{\omega}}} \right\rangle_{\rr{H}_\Gamma}\\
    &=\; \expo{-\ii \kappa\cdot \beta} \left\langle \delta_{[\kappa]}, \delta_{[\kappa+\alpha]}\right\rangle_{\ell^2(\n{B}_\Gamma)} \, \left\langle f_{\widetilde{\omega}} , F_{\gamma'_{\kappa+\alpha}} S_{\beta} f_{\widetilde{\omega}}\right\rangle_{\rr{h}_\Gamma}\\
    &= \; \chi_{\Gamma'}(\alpha)  \expo{-\ii \kappa\cdot \beta} \left\langle f_{\widetilde{\omega}}, F_{\gamma'_{\alpha}} S_{\beta}f_{\widetilde{\omega}} \right\rangle_{\rr{h}_\Gamma}\\
    &=\; \chi_{\Gamma'}(\alpha)  \expo{-\ii \kappa\cdot\beta} \left\langle f_{\widetilde{\omega}} , F_{\alpha} S_{\beta} f_{\widetilde{\omega}} \right\rangle_{\rr{h}_\Gamma}.
\end{align*}
Choosing $P:=\ketbra{f_{\widetilde{\omega}}}{f_{\widetilde{\omega}}}$, the projection over the span of $f_{\widetilde{\omega}}$, the formula holds. 
To prove that a state of this form is pure, we once again employ Proposition \ref{prop:GNS_bloch_wave_sts}. First of all, note that the dependence of the GNS triplet $(\rho,\rr{H}_\Gamma, f_{\widetilde{\omega}})$ 
on the particular state $\omega\in \s{P}_\bb{W}^{\Gamma,\beta}$ is entirely encoded in the cyclic vector $\Psi_{\kappa,0}^{f_{\widetilde{\omega}}}$
through its quasi-momentum $\kappa$ and its restriction $\widetilde{\omega}$.
 Moreover, $\rho$ is irreducible (it accommodates the GNS representation of pure states) and
 any vector of the form $\Psi_{\kappa,0}^f$, with $f\in\rr{h}_\Gamma$, is cyclic for $\rho$. 
Let $\omega_{(\kappa,P)}$ be a state of the form \ref{eq_st_pur_gamm_bl}, with $f$ some normalized vector in the image of $P$. Its GNS triplet is given by $(\rho,\rr{H}_\Gamma,\Psi_{\kappa,0}^f)$ by construction. Since $\rho$ is irreducible, we conclude that $\omega_{(\kappa,P)}$ is pure.
\qed

\begin{remark}[Explicit formulas]
The explicit form of \eqref {eq_st_pur_gamm_bl} is
\[
\omega_{(\kappa,P)}(u_\alpha v_\beta) \;:=\; \chi_{\Gamma'}(\alpha)\; \expo{-\ii \kappa\cdot \beta }\int_{\n{T}_\Gamma}\dd\nu(y)\; \expo{\ii\alpha\cdot y}\overline{f(y)}f(y-\beta)
\]
where $f\in \rr{h}_\Gamma$ is any normalized function which generates the one-dimensional subspace associated with $P$. 
This formula compares directly with the description given in \cite[Section 3]{acerbi-94}.
On the other hand by using the Fourier expansion of $f$ given by
\[
f(y)\;=\;\sum_{\gamma'\in\Gamma'}\widehat{f}(\gamma')\expo{\ii \gamma'\cdot y}
\]
one obtains
\[
\omega_{(\kappa,P)}(u_\alpha v_\beta) \;:=\; \chi_{\Gamma'}(\alpha)\; \expo{-\ii \kappa\cdot \beta}\sum_{\gamma'\in\Gamma'}\expo{-\ii \gamma'\cdot \beta}\overline{\widehat{f}(\gamma'+\alpha)}\widehat{f}(\gamma').
\]
This formula is equivalent to the one given in \cite[Theorem 3.13]{beaume-manuceau-pellet-sirugue-74}.  
 \hfill $\blacktriangleleft$
\end{remark}

\medskip

The result in Proposition \ref{cor_pas_bet-Gamma} doesn't imply that the space $\s{P}_\bb{W}^{\Gamma,\beta}$ is parametrized by the product $\n{B}_{\Gamma} \times \bb{G}_1(\rr{h}_\Gamma)$. In fact, it is also evident that the formula \eqref{eq_st_pur_gamm_bl} is not periodic in $\kappa$. A precise topological characterization of   
$\s{P}_\bb{W}^{\Gamma,\beta}$ will be presented in Section \ref{sec:classif_sts_latt}.

\subsection{Time-reversal invariant states}
Consider the morphism $\rr{c}:\bb{W}\to\bb{W}$ initially defined on the generators by
\begin{equation}\label{eq:trs}
\rr{c}(u_\alpha)\;:=\;u_\alpha^*\;=\;u_{-\alpha}\;,\qquad \rr{c}(v_\beta)\;:=\;v_\beta\;.
\end{equation}
To be compatible with the Weyl relations $\rr{c}$ must be anti-linear. 
\begin{definition}[Time-reversal symmetry]
The map $\rr{c}:\bb{W}\to\bb{W}$ initially defined by the relations \eqref{eq:trs} and then extended anti-linearly to the whole Weyl $C^*$-algebra defines an anti-linear $\ast$-involution of $\bb{W}$ called \emph{time-reversal {symmetry}}. A state $\omega\in\s{S}_{\bb{W}}$ is called \emph{time-reversal invariant} if $\overline{\omega(a)}=\omega(\rr{c}(a))$ for every $a\in\bb{W}$.
\end{definition}

\section{Topology of translationally invariant states}\label{sec:classif_sts_tras}

In this section we want to investigate the topology of  translationally invariant states.
Proposition \ref{cor:trasnl_st} establishes  the homeomorphism of topological spaces 
\[
\s{S}_{\bb{W}}^\tau\;\simeq\;\s{M}_{+,1}(\rr{b}(\n{R}^d))\;.
\]
Moreover, if one restricts to states regular in $\beta$, 
\[
 \s{S}_{\bb{W}}^{\tau,{\beta}}\;:=\;\left\{ \omega\in\s{S}_{\bb{W}}^\tau\;|\; \omega\;\;\text{is semi-regular in the parameter}\;\;\beta\right\}\;,
\]
one ends up with
\[
 \s{S}_{\bb{W}}^{\tau,{\beta}}\;\simeq\;\s{M}_{+,1}(\R^d_{\rr{b}})
 \;\simeq\;\s{M}_{+,1}(\R^d)\;,
\]
where by $\R^d_\rr{b}$ we denote $\R^d$ with the Bohr topology
and the second homeomorphism is a consequence of  Lemma \ref{lemma:regular_measure_embed}.
Since both $\s{M}_{+,1}(\rr{b}(\n{R}^d))$ and $\s{M}_{+,1}(\R^d)$ are convex sets they are in particular path connected. This implies that
\[
{\pi_0}(\s{S}_{\bb{W}}^\tau)\;=\;{\pi_0}(\s{S}_{\bb{W}}^{\tau,{\beta}})\;=\;\{[\omega_0]\}
\]
are singletons. The \virg{special} representative for the equivalence class has been chosen as the \emph{zero-momentum} state defined by
\begin{equation}\label{eq:ZMS}
\omega_0(u_\alpha v_\beta)\;=\;\delta_{\alpha,0}\;,\qquad \forall\; \alpha,\beta\in\R^d\;.
\end{equation}
To sum up, if one considers the full family of translation invariant states, or even just the semi-regular ones, one obtains the topological triviality of these classes according to the notion of path-equivalence. In the following we will show that this is the case even if one restricts to smaller classes.

\subsection{Plane-wave states}
Recall that $\s{P}_\bb{W}^\tau=\s{S}_{\bb{W}}^\tau\cap \s{P}_\bb{W}$ denotes the space of translationally invariant pure states. Now let
\[
\s{P}_\bb{W}^{\tau,\beta}\;:=\;\s{S}_{\bb{W}}^{\tau,\beta}\cap \s{P}_\bb{W}^\tau
\]
be the subset of translationally invariant pure states regular in $\beta$.
In accordance with the existing literature we will refer to elements of
$\s{P}_\bb{W}^{\tau,\beta}$ as \emph{plane-wave states}.
The next result provides a characterization of these sets, and the presented formula characterizing regular states agrees with \cite[Proposition 3.5]{beaume-manuceau-pellet-sirugue-74} and  \cite[Proposition 2.1]{strocchi-15}.

\begin{proposition}\label{prop:homeo_t_inv}
One has the homeomorphism of topological spaces 
\[
\s{P}_\bb{W}^\tau\;\simeq\; \rr{b}(\n{R}^d)\;,\qquad \s{P}_\bb{W}^{\tau,\beta}\;\simeq\;\R^d_\rr{b} \;.
\]
For every $\lambda\in\rr{b}(\n{R}^d)$ the associated state $\omega_\lambda\in \s{P}_\bb{W}^\tau$ is specified by the prescription 
\[
\omega_\lambda(u_\alpha v_\beta)\;=\;\delta_{\alpha,0}\;\xi_{-\beta}(\lambda)\;.
\]
The elements of $\s{P}_\bb{W}^{\tau,\beta}$ are parametrized by $p\in\R^d$ (called momentum) according to the prescription
\[
\omega_p(u_\alpha v_\beta)\;=\;\delta_{\alpha,0} \;\expo{-\ii p\cdot\beta}\;.
\]
Finally, one has that ${\pi_0}(\s{P}_\bb{W}^{\tau,\beta})=\{[\omega_0]\}$ is a singleton.
\end{proposition}
\proof
We will prove that the map $\Psi:\lambda\mapsto \omega_\lambda$ provides an homeomorphism $\rr{b}(\R^d)\simeq\s{P}_\bb{W}^\tau$. Then the homeomorphism $\R^d_\rr{b} \simeq \s{P}_\bb{W}^{\tau,\beta}$ is obtained as a restriction and both formulas follow by Proposition \ref{cor:trasnl_st}, as pure states correspond under this identification to Dirac measures. 
First, let us observe that $\rr{b}(\n{R}^d) \simeq \s{P}_\bb{V}\simeq \s{P}_\bb{W}^\tau$. 
The first homeomorphism is given by the Gelfand isomorphism through the map $\lambda\mapsto \widetilde{\omega}_\lambda$, and the second one is the homeomorphism $\widetilde{\omega}_\lambda \mapsto \omega_\lambda := \widetilde{\omega}_\lambda \circ \langle\,\cdot\,\rangle$ in Proposition \ref{prop_inv_sts}. The  composition of these two identifications provides the homeomorphism $\Psi$.
For the last assertion note that $\R^d_\rr{b}$ is path connected as the image of $\R$ under the continuous map $\rr{i}_\rr{b}$ (\cf Theorem \ref{thm:bohr_comp_map}).
\qed

\medskip

As a consequence of the previous result it turns out that plane-wave states are topologically trivial in the sense that they are all equivalent to the \emph{zero momentum state} $\omega_0$. This state has a special role inside $\s{P}_\bb{W}^\tau$.
\begin{proposition}
The {zero momentum state} $\omega_0$ is the unique element of $\s{P}_\bb{W}^\tau$ which is time-reversal invariant.
\end{proposition}
\proof
In view of Proposition \ref{prop:homeo_t_inv} the condition $\omega_\lambda\circ \rr{c}=\overline{\omega_\lambda}$ translates into $\xi_\beta(\lambda)=\overline{\xi_\beta(\lambda)}=\xi_{-\beta}(\lambda)$ for every $\beta\in\R^d$.
Using the fact that the $\xi_\beta$ are characters of  $\rr{b}(\n{R}^d)$ one obtains $1=\xi_0(\lambda)=\xi_{\beta-\beta}(\lambda)=\xi_\beta(\lambda)^2$ and in turn $\xi_\beta(\lambda)\in\{\pm 1\}$ for every $\beta\in\R^d$.
Now suppose there exists some $\sigma\in\R^d$ such that $\xi_\sigma(\lambda) = -1$. Then, $\xi_{\sigma/2}(\lambda)^2 = \xi_\sigma(\lambda) = -1$, so $\xi_{\sigma/2}(\lambda)\in\{\pm \ii\}$, contradicting the above. Hence, $\xi_\beta(\lambda) = 1$ for all $\beta\in\R^d$, which implies $\lambda=0$.
\qed

The fact that every plane-wave state is equivalent to $\omega_0$ also holds true under the stronger notion of automorphic equivalence.

\begin{proposition}\label{prop_aut_eq_planewave}
    Every element of $\s{P}_\bb{W}^{\tau,\beta}$ is automorphically equivalent to $\omega_0$.
\end{proposition}
\proof
For $\alpha\in\R^d$, define an automorphism $\theta_\alpha\in{\rm Aut}(\bb{W})$ by $\theta_\alpha(a) = u_\alpha a u_\alpha^*$, for any $a\in\bb{W}$. By analogy with $\tau_\lambda$ in Proposition \ref{prop:transl_str_cont}, this is a strongly continuous family of automorphisms, called \textit{momentum translations}. Note that $\omega_p\circ\theta_\alpha = \omega_{p-\alpha}$ by a quick check in the generators. Now, for any $\omega_p\in \s{P}_\bb{W}^{\tau,\beta}$, with momentum $p$, define the function $[0,1]\ni t\mapsto \lambda_{t}:= \theta_{tp}\in{\rm Aut}(\bb{W})$, which is strongly continuous by the above. Then, it holds that $t\mapsto \omega_p\circ \lambda_{t} = \omega_{p(1-t)}$ is a continuous path through $\s{P}_\bb{W}^{\tau,\beta}$, given by automorphisms, which joins $\omega_p$ and $\omega_0$.
\qed

\medskip

\begin{remark}[Topology of irregular pure states]\label{rk:irreg_phas}
Proposition \ref{prop:homeo_t_inv} only asserts that all the pure and $\beta$-regular states are equivalent to the zero-momentum state. For the case of the irregular states, Proposition \ref{prop:homeo_t_inv}
asserts that ${\pi_0}(\s{P}_\bb{W}^\tau)$ corresponds with the path components of $\rr{b}(\R^d)$. Interestingly, the latter space is not path-connected as proved in Proposition \ref{prop:conex_bohr}, which means that there are classes of irregular states which are not path-equivalent to the zero-momentum state. 
Thus, non-trivial topological phases are bound to emerge if one considers configurations valued in both regular and irregular invariant pure states. 
Due to the pathological nature of the elements of $\rr{b}(\R^d)\setminus\R^d_{\rr{b}}$, corresponding to discontinuous, hence non-measurable characters of $\R^d$, we find that this result is physically reasonable. This is in view of the fact that in such states, momentum cannot be employed as a quantum number. In fact, let $\omega\in\s{P}_\bb{W}^\tau\setminus \s{P}_\bb{W}^{\tau,\beta}$ be such a state, and $(\pi_\omega,\s{H}_\omega,\psi_\omega)$ its GNS representation. Then, since the unitary group representation $\beta\mapsto\pi_\omega(v_\beta)$ is not strongly continuous, one cannot define a momentum operator as a strong derivative (\ie, as its infinitesimal generator), as is usually done when using Weyl quantization.
\hfill $\blacktriangleleft$
\end{remark}

\section{Topology of pure lattice-invariant states}\label{sec:classif_sts_latt}

Even though Proposition \ref{cor_pas_bet-Gamma} states that the elements $\s{P}_\bb{W}^{\Gamma,\beta}$ are parametrized 
by the  points of the fundamental cell $Q_{\Gamma'}$, and the latter is in bijection with the torus $\n{B}_\Gamma$, 
it is not  immediately evident that the  elements $\s{P}_\bb{W}^{\Gamma,\beta}$ can be parametrized by points in $\n{B}_\Gamma$. The reason being formula \eqref{eq_st_pur_gamm_bl}
is not $\Gamma'$-periodic in $\kappa$. In fact, for every $\gamma'\in\Gamma'$ one gets
\[
\omega_{(\kappa+\gamma',P)}(u_\alpha v_\beta)\;=\;\expo{-\ii \gamma'\cdot\beta}\omega_{(\kappa,P)}(u_\alpha v_\beta)
\;\neq\;\omega_{(\kappa,P)}(u_\alpha v_\beta)\;.
\]
However, if one introduces the family of 
automorphisms
\begin{equation}\label{eq_lam_aut}
\lambda_{\gamma'}(A)\;:=\;F_{\gamma'}AF_{\gamma'}^*\;,\qquad A\in\bb{B}(\rr{h}_\Gamma) 
\end{equation}
 one gets that
\begin{equation}\label{eq:cov_st}
\omega_{(\kappa+\gamma',P)} \;=\;\omega_{(\kappa,\lambda_{\gamma'}(P))}
\end{equation}
for every  $\gamma'\in\Gamma'$. We will refer to \eqref{eq:cov_st} as the \emph{covariance property}. This will be a key ingredient for the analysis of the topological structure of the set $\s{P}_\bb{W}^{\Gamma,\beta}$ described in the next section.

\subsection{Bloch-wave states}\label{Se:BW}
Let us start by observing that the set $\s{P}_\bb{W}^{\Gamma,\beta}$ cannot be naively identified with the Cartesian product $\n{B}_{\Gamma} \times \bb{G}_1(\rr{h}_\Gamma)$ in view of the {covariance property}
\eqref{eq:cov_st}. To accommodate this let us consider on the product space 
$\R^d\times \rr{h}_\Gamma$ the $\Gamma'$-action defined by $\gamma':(x,f)\to(x+\gamma',F_{\gamma'}f)$. As explained in Appendix \ref{app:G-bund}
the quotient space 
\[
\bb{E}_\Gamma\;:=\;\big(\R^d \times  \rr{h}_\Gamma\big)/\Gamma'
\]
inherits the structure of a Hilbert bundle over the base space $\n{B}_\Gamma$ with typical fiber $\rr{h}_\Gamma$.
Since ${\rm dim}(\rr{h}_\Gamma)=\aleph_0$, in view of the argument at the end of  Appendix \ref{app:hilb_bun}, it turns out that $\bb{E}_\Gamma \simeq \n{B}_\Gamma\times \rr{h}_\Gamma$ is isomorphic to the trivial bundle. 
Let ${^{\rm w}\bb{G}_1}(\rr{h}_\Gamma)$ be the space $\bb{G}_1(\rr{h}_\Gamma)$  endowed with the weak topology. In a similar way the product space
$\R^d\times {^{\rm w}\bb{G}_1}(\rr{h}_\Gamma)$ can be endowed with the $\Gamma'$-action
 $\gamma':(x,P)\mapsto(x+\gamma',\lambda_{-\gamma'}(P))$.  Then
\begin{equation}\label{eq:G-obun_k1}
\rr{Gr}_1(\bb{E}_\Gamma)\;:=\;\big(\R^d\times {^{\rm w}\bb{G}_1}(\rr{h}_\Gamma)\big)/\Gamma'
\end{equation}
turns out to be the Grassmann bundle of rank $k=1$ associated with the Hilbert bundle $\bb{E}_\Gamma$, equipped with the weak topology in its fibers. As justified at the end of Appendix \ref{app:G-bund}, it turns out that 
$\rr{Gr}_1(\bb{E}_\Gamma)\simeq \n{B}_\Gamma\times {^{\rm w}\bb{G}_1}(\rr{h}_\Gamma)$
is indeed trivial. We are now in position to prove a key result of this work.
\begin{theorem}\label{thm:homeo_gamma_inv}
The prescription \eqref{eq_st_pur_gamm_bl} provides an homeomorphism
\[
\s{P}_\bb{W}^{\Gamma,\beta}\;\simeq\;\rr{Gr}_1(\bb{E}_\Gamma) \;.
\]
As a consequence ${\pi_0}(\s{P}_\bb{W}^{\Gamma,\beta})=\{[\omega_0]\}$ {is a singleton}.
\end{theorem}
\proof
Let us introduce the map $\Phi:\R^d\times {^{\rm w}\bb{G}_1}(\rr{h}_\Gamma)\to \s{P}_\bb{W}^{\Gamma,\beta}$ defined by $\Phi:(x,P)\mapsto \omega_{(x,P)}$, where $\omega_{(x,P)}$ is given by the formula \eqref{eq_st_pur_gamm_bl} (tacitly extended to all $x\in\R^d$).
In view of the covariance property \eqref{eq:cov_st}, this map is $\Gamma'$-periodic in the sense
\[
\Phi(x+\gamma',\lambda_{-\gamma'}(P))\;=\; \Phi(x,P)
\]
for every $\gamma'\in\Gamma'$. Let us show that it is continuous. Since the space $\R^d\times {^{\rm w}\bb{G}_1}(\rr{h}_\Gamma)$ is separable and metrizable in view of \cite[Corollary 2.5]{shubin-96} one can check continuity on sequences. Let $\{(x_n,P_n)\}_{n\in\N}$ be a sequence in $\R^d\times {^{\rm w}\bb{G}_1}(\rr{h}_\Gamma)$ such that $(x_n,P_n)\to(x,P)$. In order to show that  $\Phi(x_n,P_n)\to\Phi(x,P)$, one must prove that $\omega_{(x_n,P_n)}\to \omega_{(x,P)}$ in the $\ast$-weak topology. For that, it is enough to show that
\begin{equation}\label{eq:con_equi}
\lim_{n\to\infty}\left|\omega_{(x_n,P_n)}(u_\alpha v_\beta)- \omega_{(x,P)}(u_\alpha v_\beta)\right|\;=\;0
\end{equation}   
for every $\alpha, \beta\in\R^d$. Indeed from \eqref{eq:con_equi} it follows by linearity that $\omega_{(x_n,P_n)}(a_0)\to \omega_{(x,P)}(a_0)$ for every $a_0\in\bb{W}_0$, and by the density of $\bb{W}_0$ in $\bb{W}$ and the inequality
\begin{align*}
    |\omega_{(x_n,P_n)}(a)-\omega_{(\lambda,P)}(a)| &\leq |\omega_{(x_n,P_n)}(a-a_0)| + |\omega_{(x_n,P_n)}(a_0)-\omega_{(\lambda,P)}(a_0)| \\ &+ |\omega_{(\lambda,P)}(a-a_0)|\;,
\end{align*}
this implies that $\omega_{(x_n,P_n)}(a)\to \omega_{(x,P)}(a)$ for every $a\in\bb{W}$.
To check \eqref{eq:con_equi}, it is enough to use the prescription \eqref{eq_st_pur_gamm_bl} along with the observations that $|\expo{\ii x_n\cdot \beta}-\expo{\ii x\cdot \beta}|\to 0$ if $x_n\to x$ and 
\[
\begin{aligned}
\left|{\rm Tr}_{\rr{h}_\Gamma}(P_nF_{\alpha}S_\beta)-{\rm Tr}_{\rr{h}_\Gamma}(PF_{\alpha}S_\beta)\right|\;&=\; \left|{\rm Tr}_{\rr{h}_\Gamma}((P_n-P)F_{\alpha}S_\beta)\right|\\
&\leqslant\;\|(P_n-P)F_{\alpha}S_\beta\|_1\;\longrightarrow\;0
\end{aligned}
\]
when $P_n\to P$ in the weak topology. The last fact follows immediately from \cite[Theorem 2.20]{simon-05} and $\|\cdot\|_1$ denotes the norm of the Schatten ideal $\bb{L}^1(\rr{h}_\Gamma)$ of trace class operators. In fact, since $F_{\alpha}S_\beta$ are unitary operators one gets that 
\[
|QF_{\alpha}S_\beta|\;=\;S_\beta^*F_{\alpha}^* Q F_{\alpha}S_\beta\;,\quad |(QF_{\alpha}S_\beta)^*|\;=\;Q
\]
where $Q$ stands both for $P_n$ or $P$. From this it follows immediately that $\|QF_{\alpha}S_\beta\|_1=1$ and all the conditions of \cite[Theorem 2.20]{simon-05} are satisfied. Since the map $\Phi$ is $\Gamma'$-periodic and continuous, it factors through the quotient to a continuous map (still denoted with the same symbol) $\Phi:\rr{Gr}_1(\bb{E}_\Gamma)\to\s{P}_\bb{W}^{\Gamma,\beta}$. This map is bijective in view of Proposition \ref{cor_pas_bet-Gamma}. 
Now, note ${^{\rm w}\bb{G}_1}(\rr{h}_\Gamma)$ is path-connected, since the identity map ${^{\rm u}\bb{G}_1}(\rr{h}_\Gamma)\to {^{\rm w}\bb{G}_1}(\rr{h}_\Gamma)$, where ${^{\rm u}\bb{G}_1}(\rr{h}_\Gamma)$ is the same space equipped with the uniform topology, is continuous, and ${^{\rm u}\bb{G}_1}(\rr{h}_\Gamma)$ is path-connected \cite[Section 2.3]{abbondandolo-majer-09}. Being $\Phi$ a continuous bijection, the last assertion follows since $\rr{Gr}_1(\bb{E}_\Gamma)$ is path-connected, as a quotient of a product of path-connected spaces.
The last task is to prove that $\Phi$  is an homeomorphism.
From \cite[Proposition 3.5]{shubin-96}, one knows that the weak closure of $\bb{G}_1(\rr{h}_\Gamma)$ in $\bb{B}(\rr{h}_\Gamma)$ is
\[
\overline{{^{\rm w}\bb{G}_1}(\rr{h}_\Gamma)}\;=\;\{ aP \; |\; a\in[0,1]\;, P\in {^{\rm w}\bb{G}_1}(\rr{h}_\Gamma)\}\;\subset\;\bb{J}_{[0,1]}
\]
where $\bb{J}_{[0,1]}\subset \bb{B}(\rr{h}_\Gamma)$ is  the subset of self-adjoint operators $A$ such that $0 \leq A\leq {\bf 1}$. The compactness of $\bb{J}_{[0,1]}$ in the weak topology \cite[p. 196]{shubin-96} implies the compactness of $\overline{{^{\rm w}\bb{G}_1}(\rr{h}_\Gamma)}$. The automorphisms \eqref{eq_lam_aut} fix $\overline{{^{\rm w}\bb{G}_1}(\rr{h}_\Gamma)}$ and one can define the quotient space
\[
\rr{Tr}_1(\bb{E}_\Gamma)\; :=\; \big(\R^d\times \overline{{^{\rm w}\bb{G}_1}(\rr{h}_\Gamma)}\big)/\Gamma'
\;\simeq\;\n{B}_\Gamma \times \overline{{^{\rm w}\bb{G}_1}(\rr{h}_\Gamma)}
\]
which contains $\rr{Gr}_1(\bb{E}_\Gamma)$. 
The second homeomorphism, which is a consequence of the triviality of  $\rr{Gr}_1(\bb{E}_\Gamma)$, implies that $\rr{Tr}_1(\bb{E}_\Gamma)$ is compact. 
Let  ${^{\rm w}}\bb{W}^*$ be the dual space of $\bb{W}$ equipped with the weak-$\ast$ topology and define the map $\widehat{\Phi}: \rr{Tr}_1(\bb{E}_\Gamma)\to {^{\rm w}}\bb{W}^*$, by $\widehat{\Phi}([\kappa,T]) = \omega_{[\kappa,T]}$, with
\[
\omega_{[\kappa,T]}(u_\alpha v_\beta) \;:=\; \chi_{\Gamma'}(\alpha)\; \expo{-\ii \kappa\cdot \beta }{\rm Tr}_{\rr{h}_\Gamma}(T F_{\alpha}S_\beta)\,.
\]
Clearly, $\widehat{\Phi}$ is an extension of $\Phi$. It is also continuous, as can be checked by an argument analogous to the one used for $\Phi$. Moreover, by writing $T = a_T P_T$, with $a_T\in[0,1]$ and $P_T\in {^{\rm w}\bb{G}_1}(\rr{h}_\Gamma)$ (the spectral decomposition of $T$), one gets
\[
\widehat{\Phi}([\kappa,T]) \;=\; \omega_{[\kappa,T]} \;=\; \omega_{[\kappa,a_T\,P_T]} \;=\; a_T\, \omega_{[\kappa,P_T]} \;=\; a_T\, \Phi([\kappa,P_T]).
\]
With this, suppose $\widehat{\Phi}([\kappa,T]) = \widehat{\Phi}([\kappa',T'])$ for some $\kappa,\kappa'\in\n{B}_\Gamma$, $T,T'\in \overline{{^{\rm w}\bb{G}_1}(\rr{h}_\Gamma)}$. Then,
\begin{align*}
    a_T\, \Phi([\kappa,P_T]) \;=\; a_{T'}\, \Phi([\kappa',P_{T'}]),
\end{align*}
which by evaluating both sides in ${\bf 1}$ implies $a_T = a_{T'}$, and then by injectivity of $\Phi$, $[\kappa,P_T] = [\kappa',P_{T'}]$. Combining both equations, we get $[\kappa,T] = [\kappa',T']$, and thus $\widehat{\Phi}$ is injective. Therefore, since its domain is compact and its codomain is Hausdorff, $\widehat{\Phi}$ is a topological embedding and, consequently, $\Phi$ is an homeomorphism as a bijective restriction of $\widehat{\Phi}$.
\qed

\medskip

In order to complete the description of Theorem \ref{thm:homeo_gamma_inv}, let us 
note that the zero-momentum state $\omega_0$ is indeed a Bloch-wave state, and thus a candidate for the representative in the description of ${\pi_0}(\s{P}_\bb{W}^{\Gamma,\beta})$. This is because $\s{P}_\bb{W}^{\tau,\beta}\subset \s{P}_\bb{W}^{\Gamma,\beta}$, since translation-invariant states are in particular $\Gamma$-invariant for every lattice $\Gamma$.
In terms of the prescription  \eqref{eq_st_pur_gamm_bl}, one can check that $\omega_0$ corresponds to the state associated with the pair $(0,P_0)$ where $P_0:=\ketbra{\xi_0}{\xi_0}$ is the rank 1 projection on the (constant) state $\xi_0(y):=1$ of $\rr{h}_\Gamma$.

\medskip

It is worth characterizing the elements of $\s{P}_\bb{W}^{\Gamma,\beta}$ that are invariant under time-reversal symmetry. For that, 
let us introduce some notation. Let $C$ be the operator which implements the complex conjugation on $\rr{h}_\Gamma$, \ie $Cf:=\overline{f}$
for every $f\in \rr{h}_\Gamma$.
A direct check shows   $CF_\alpha S_\beta C = F_{-\alpha} S_\beta$, for any $\alpha\in\Gamma'$ and $\beta\in\R^d$.
Let $\n{B}_\Gamma^c$ be the set of  points of $\n{B}_\Gamma$  fixed by reflection, \ie such that $\kappa=-\kappa$.
This set is represented by the elements of $Q_{\Gamma'}$ such that  
$2\kappa\in \Gamma'$. Therefore the cardinality of $\n{B}_\Gamma^c$ is  exactly $2^d$, and the representatives of these points are the   
linear combinations with coefficients $0$ or ${1}/{2}$ of the basis vectors $\{\rr{b}^1,\dots,\rr{b}^d\}$ of $\Gamma'$.
\begin{proposition}\label{prop:blochwave_sts_TRI}
     The element $\omega_{(\kappa,P)} \in \s{P}_\bb{W}^{\Gamma,\beta}$ is time-reversal invariant if and only if $\kappa\in \n{B}_\Gamma^c$  and $P$ satisfies $CPC = \lambda_{2\kappa} (P)$.
\end{proposition}
\proof 
Let us compute $\overline{\omega_{(\kappa,P)}\circ\rr{c}(u_{\alpha}v_\beta)} = \overline{\omega_{(\kappa,P)}(u_{-\alpha}v_\beta)}$. One has,
\[
\overline{\omega_{(\kappa,P)}(u_{-\alpha}v_\beta)} \;=\; \chi_{\Gamma'}(\alpha) \expo{\ii\kappa\cdot\beta} \overline{{\rm Tr}_{\rr{h}_\Gamma}(PF_{-\alpha} S_\beta)}.
\]
Now, since for any $A\in\bb{B}(\rr{h}_\Gamma)$ it holds true that ${\rm Tr}_{\rr{h}_\Gamma}(CAC) = \overline{{\rm Tr}_{\rr{h}_\Gamma}(A)}$, one gets
\begin{align*}
    \overline{\omega_{(\kappa,P)}(u_{-\alpha}v_\beta)} \;&=\; \chi_{\Gamma'}(\alpha) \expo{\ii\kappa\cdot\beta} {\rm Tr}_{\rr{h}_\Gamma}(CPF_{-\alpha} S_\beta C) \\
    &=\; \chi_{\Gamma'}(\alpha) \expo{\ii\kappa\cdot\beta} {\rm Tr}_{\rr{h}_\Gamma}(CPC F_{\alpha}  S_\beta)\\
    &=\; \omega_{(-\kappa,CPC)}(u_\alpha v_\beta)\;.
\end{align*}
 From this, we conclude the equation  
\[
\overline{\omega_{(\kappa,P)}\circ\rr{c}} \;=\; \omega_{(-\kappa,CPC)} \;.
\]
Thus $\omega_{(\kappa,P)}$ is time-reversal invariant if and only if $\omega_{(\kappa,P)} = \omega_{(-\kappa,CPC)}$, which by Theorem \ref{thm:homeo_gamma_inv} happens if and only if there exists $\gamma'\in\Gamma'$ such that $\kappa = -\kappa+\gamma'$ and $P = \lambda_{-\gamma'}(CPC)$. This means that $\kappa\in \n{B}_\Gamma^c$  and $\lambda_{2\kappa}(P) = CPC$, where we are tacitly using $\kappa$  for denoting the representative in the unit cell.
\qed

\medskip

\section{Topology of Zak states}\label{app-zac}
We will consider here a subclass of $\Gamma$-invariant states that are strictly more symmetric, being also invariant under the momentum translations by the dual lattice $\Gamma'$.
Let us recall the notation $\theta_{\alpha}(a):=u_\alpha a u_\alpha^*$ for every $a\in\bb{W}$ and $\alpha\in\R^d$ for the 
momentum translations as introduced in Lemma \ref{lemma:GNS_quasimomentum_kappa}.

\begin{definition}[Zak states]
    A state $\omega\in\s{S}_\bb{W}$ such that $\omega\circ \tau_{\gamma} \circ \theta_{\gamma'}   = \omega$ for all $(\gamma,\gamma')\in\Gamma\times \Gamma'$
        is said to be a \emph{Zak state}. The set of Zak states will be denoted with $\s{S}_{\bb{W}}^{\rm Z}$, and the set of pure Zak states with $\s{P}_{\bb{W}}^{\rm Z}$.
\end{definition}

In a completely analogous way to $\Gamma$-invariant states, we may define the algebra of invariant elements under $\Gamma\times\Gamma'$ as
\[
{\rm Inv}_{\tau,\theta}^{\Gamma\times\Gamma'}(\bb{W})\;:=\;\{a\in\bb{W}\;|\;  \tau_{\gamma} \circ \theta_{\gamma'}(a)=a\;,\;\;\forall\; (\gamma,\gamma')\in\Gamma\times\Gamma'\}\;,
\]
and the ergodic mean
\begin{equation}\label{eq:er_mean_zak}
\langle a\rangle_{\Gamma\times\Gamma'} \;:=\; \lim_{N\to+\infty}\frac{1}{|\Gamma_N|^2} \sum_{(\gamma,\gamma')\in\Gamma_N\times\Gamma'_N}  \tau_{\gamma} \circ \theta_{\gamma'}(a)
\end{equation} 
where $N\in\N$,  $\Gamma_N:=\Gamma\cap\Lambda_N$, $\Gamma'_N:=\Gamma'\cap\Lambda_N$ and $|\Gamma_N| = |\Gamma'_N| =(2N+1)^d$.

\medskip

Let $\bb{Z}_\Gamma$ be the $C^*$ algebra generated by the elements $u_{\gamma'} v_\gamma$, with $(\gamma,\gamma')\in\Gamma\times\Gamma'$. Just like $\bb{V}$, this is a maximal commutative $C^*$-subalgebra of $\bb{W}$, and can be proven to be isomorphic to the group $C^*$ algebra $C^*(\Gamma\times\Gamma')$ via the map $u_{\gamma'}v_\gamma \mapsto \delta_{(\gamma,\gamma')}$. From this relation, one obtains that
\begin{equation}\label{zak_spectrum}
    \bb{Z}_\Gamma \;\simeq\; C(\n{B}_\Gamma \times \n{T}_\Gamma)
\end{equation}
with the isomorphism  implemented by the Gelfand transform $u_{\gamma'} v_\gamma \mapsto \zeta_{(\gamma,\gamma')}$, where $\zeta_{(\gamma,\gamma')}(\kappa,\nu) = \expo{-\ii (\kappa\cdot \gamma + \nu\cdot \gamma')}$. The next results are analogous to Propositions \ref{prop_inv_tau_charac} and \ref{prop_gamma_inv_sts}, and to Propositions \ref{prop_inv_sts} and \ref{prop_inv_sts_oo}, respectively.
\begin{proposition}\label{prop_zak_obs}
It holds true that ${\rm Inv}_{\tau,\theta}^{\Gamma\times\Gamma'}(\bb{W}) = \bb{Z}_\Gamma$.
\end{proposition}

\begin{proposition}\label{prop_inv_sts_zak}
There is an homeomorphism  $\s{S}_{\bb{W}}^{\rm Z} \simeq \s{S}_{\bb{Z}_\Gamma}$ given as follows: every $\widetilde{\omega}\in \s{S}_{\bb{Z}_\Gamma}$ defines a state $\omega \in \s{S}_{\bb{W}}^{\rm Z}$ through the prescription
\begin{equation}\label{eq_st_inv_zak}
\omega(a)\;:=\;\widetilde{\omega}(\langle a\rangle_{\Gamma,\Gamma'})\;,\qquad \forall\; a\in \bb{W}\;,
\end{equation}
and all the elements of $\s{S}_{\bb{W}}^{\rm Z}$ are of this form. Moreover,  this  homeomorphism restricts to an homeomorphism $\s{P}_\bb{W}^{\rm Z}\simeq \s{P}_{\bb{Z}_\Gamma}$.
\end{proposition}
\proof The proof is analogous to the one for Proposition \ref{prop_inv_sts}. In contrast with $\bb{V}_\Gamma$ in Proposition \ref{prop_inv_sts_oo}, $\bb{Z}_\Gamma$ is (maximal) commutative, so $\widetilde{\omega}\in \s{P}_{\bb{Z}_\Gamma}$ implies $\omega \in \s{P}_{\bb{W}}^{\rm Z}$
and we have the homeomorphism $\s{P}_\bb{W}^{\rm Z}\simeq \s{P}_{\bb{Z}_\Gamma}$ in this case as well.
\qed

\medskip

The next result is a direct consequence of Proposition \ref{prop_inv_sts_zak} and the isomorphism \eqref{zak_spectrum}.
\begin{proposition}[Topology of pure Zak states]\label{cor_pas_bet-Zak}
Every element of $\s{P}_{\bb{W}}^{\rm Z}$ is of the form
\begin{equation}\label{eq_st_pur_zak}
\omega_{(\kappa,\nu)}(u_\alpha v_\beta) \;:=\; \chi_{\Gamma'}(\alpha) \, \chi_{\Gamma}(\beta)\; \expo{-\ii \kappa\cdot \beta } \expo{-\ii \alpha \cdot \nu}\;
\end{equation}
where $(\kappa,\nu)\in \n{B}_\Gamma \times \n{T}_\Gamma$. This correspondence is in fact an homeomorphism.
\end{proposition}
\proof The isomorphism \eqref{zak_spectrum} implies that $\s{P}_{\bb{Z}_\Gamma}\simeq \n{B}_\Gamma \times \n{T}_\Gamma$ through the correspondence $\widetilde{\omega}_{(\kappa,\nu)}(u_{\gamma'} v_\gamma) = \zeta_{({\gamma'},\gamma)}(\kappa,\nu) = \expo{-\ii (\kappa\cdot \gamma + \nu \cdot {\gamma'})}$, where $\zeta_{({\gamma'},\gamma)}$ is the Gelfand transform of $u_{\gamma'} v_\gamma$. Then, Proposition \ref{prop_inv_sts_zak} implies the desired result.
\qed

\medskip

One may use the states of the form \eqref{eq_st_pur_zak} to reconstruct all possible Zak states.
\begin{corollary}
    There is an homeomorphism $\s{S}_{\bb{W}}^{\rm Z} \simeq \s{M}_{+,1}(\n{B}_\Gamma\times \n{T}_\Gamma)$ given as follows: for any state $\omega\in\s{S}_{\bb{W}}^{\rm Z}$ there exists a unique measure $\mu_\omega\in \s{M}_{+,1}(\n{B}_\Gamma\times \n{T}_\Gamma)$ such that
    \[
    \omega(a) \;=\; \int_{\n{B}_\Gamma\times \n{T}_\Gamma} \dd\mu_\omega(\kappa,\nu) \; \omega_{(\kappa,\nu)} (a) \;, \qquad \forall a\in \bb{W}\;.
    \]
\end{corollary}

\medskip

Formula \eqref{eq_st_pur_zak} shows that  Zak states are examples of pure and $\Gamma$-invariant irregular states of $\bb{W}$. As such, even though both $\s{P}_\bb{W}^{\Gamma,\beta}$ and $\s{P}_{\bb{W}}^{\rm Z}$ are subsets of $\s{P}_\bb{W}^{\Gamma}$, it holds true that $\s{P}_\bb{W}^{\Gamma,\beta}\cap\s{P}_{\bb{W}}^{\rm Z}=\emptyset$.

We may also classify the pure Zak states that are time-reversal invariant.
\begin{proposition}
    The state $\omega_{(\kappa,\nu)}\in \s{P}_{\bb{W}}^{\rm Z}$ is time-reversal invariant if and only if $\kappa\in \n{B}_\Gamma^c$.
\end{proposition}
\proof One has
\begin{align*}
\overline{\omega_{(\kappa,\nu)}(u_{-\alpha}v_\beta)} \;&=\; \chi_{\Gamma'}(\alpha) \, \chi_{\Gamma}(\beta)\; \overline{\expo{-\ii \kappa\cdot \beta } \expo{+\ii \alpha \cdot \nu}}\\
&=\;\chi_{\Gamma}(\beta)\; \expo{+\ii \kappa\cdot \beta } \expo{-\ii \alpha \cdot \nu} \;=\; \omega_{(-\kappa,\nu)}(u_{\alpha}v_\beta)\;,
\end{align*}
from which the claim easily follows.
\qed

Zak states are also all automorphically equivalent.
\begin{proposition}
    Every element of $\s{P}_\bb{W}^{\rm Z}$ is automorphically equivalent to $\omega_{(0,0)}$.
\end{proposition}
\proof
The proof is analogous to that of \ref{prop_aut_eq_planewave}, now using $\tau_\beta\circ\theta_\alpha$ instead of just $\theta_\alpha$ as the base for the family of automorphisms.
\qed

\appendix

\section{A  toolkit in bundle theory}
\label{sec:topol}

During this section $\rr{h}$ will denote a separable complex Hilbert space of dimension ${\rm dim}(\rr{h})=\aleph_0$ and ${\rm Tr}_{\rr{h}}$ the canonical trace on it. The $C^*$-algebra of bounded operators on $\rr{h}$ will be denoted with $\bb{B}(\rr{h})$, the group of unitary operators by $\bb{U}(\rr{h})$, and the ideal of trace-class operators with $\bb{L}^1(\rr{h})$. An orthogonal projection is an element $P\in\bb{B}(\rr{h})$ such that $P=P^*=P^2$. The set of all projections of $\rr{h}$ will be denoted with $\bb{P}(\rr{h})$.
A projection has finite rank $k\in\N$ when ${\rm Tr}_{\rr{h}}(P)=k$.
In the following we will denote with $X$ a Hausdorff space
with the homotopy type of a CW-complex. The latter condition implies that $X$ is paracompact.

\subsection{Hilbert bundles}\label{app:hilb_bun}
The content of  this section is mainly based on \cite{dupre-74,espinoza-uribe-14,schottenloher-18}. 

\medskip

A \emph{Hilbert bundle} $\bb{H}$ over the (\emph{base}) space $X$ with \emph{typical fiber} $\rr{h}$ is a \emph{locally trivial} fiber bundle
$\pi:\bb{H}\to X$ such that: (i) the projection $\pi$ is continuous and surjective (hence open); (ii) for every $x\in X$ the fiber $\bb{H}_x:=\pi^{-1}(x)$ is \emph{unitarily} isomorphic to $\rr{h}$. One usually refers to $\bb{H}$ as the \emph{total} space of the   bundle. 
Local triviality means that there exists a cover of open subsets $V\subset X$ with bundle charts (\ie homeomorphisms) $\phi:\bb{H}|_V\to V\times\rr{h}$
such that ${\rm pr}_1\circ \phi=\pi$ and
\[
\phi_x\;:=\;{\rm pr}_2\circ \phi|_{\bb{H}_x}\;:\;\bb{H}_x\;\rightarrow\;\rr{h}
\]
is unitary for all $x\in X$. Here ${\rm pr}_1$ and ${\rm pr}_2$ are the canonical projections from $V\times\rr{h}$ to $V$ and $\rr{h}$, respectively. 
The transition map for another bundle chart $\phi':\bb{H}|_{V'}\to {V'}\times\rr{h}$ such that $W:=V\cap V'\neq\emptyset$ is given by $\phi'\circ \phi^{-1}: W\times\rr{h}\to W\times\rr{h}$, and is completely determined by the projection
\[
\psi_{V,V'}\;:=\;{\rm pr}_2\circ \left( \phi'\circ \phi^{-1}\right)\;:\;W\times\rr{h}\;\rightarrow\; \rr{h}\;.
\]
Let $G_{V,V'}:W\to {^{\rm w}}\bb{U}(\rr{h})$ be the map defined as $G_{V,V'}(x)f:=\psi_{V,V'}(x,f)$ for every $x\in W$ and $f\in \rr{h}$.
The maps $G_{V,V'}$ provide the (continuous) transition functions between the open subsets $V$ and $V'$, therefore ${^{\rm w}}\bb{U}(\rr{h})$ is the \emph{structural group} of the Hilbert bundles. 
 
\medskip

The simplest example of a Hilbert bundle is the \emph{trivial} (or product) Hilbert bundle ${\rm pr}_1:\bb{H}_0:=X\times \rr{h}\to X$. Two Hilbert bundles $\pi:\bb{H}\to X$ and $\pi':\bb{H}'\to X$ are isomorphic if there exists a continuous map $\eta:\bb{H}\to\bb{H}'$ such that $\pi'=\pi\circ \eta$ and the fiber restrictions $\eta_x:\bb{H}_x\to \bb{H}_x'$ are unitary operators for every point $x\in X$. The assignation of the transition maps $x\mapsto \eta_x$ is continuous in $^{\rm w}\bb{U}(\rr{h})$. We will denote with ${\rm Hilb}_{\rr{h}}(X)$ the set of isomorphism classes of Hilbert bundles over $X$ with typical fiber $\rr{h}$.
The classification of Hilbert bundles can be described by using the  \emph{classifying space} $B({^{\rm w}}\bb{U}(\rr{h}))$.
Recall that a classifying space $B\n{G}$ of a topological group $\n{G}$ can be realized (through the Milnor construction) as the quotient of a weakly contractible space $E\n{G}$ (\ie a topological space for which all its homotopy groups are trivial) by a proper free action of $\n{G}$.
By definition $B\n{G}$ is unique up to weak homotopy equivalence. 
Let us denote with $[X,B\n{G}]$ the set of homotopy classes of continuous maps from the topological space $X$ to $B\n{G}$. Observe that in the case $X$ is (homotopy equivalent to) a CW-complex, $[X,B\n{G}]$ doesn't change under weak homotopy equivalences of the classifying space $B\n{G}$. 
\medskip

One has the following result \cite{dupre-74,schottenloher-18}. Since $^{\rm w}\bb{U}(\rr{h})$ is contractible, its classifying space $B(^{\rm w}\bb{U}(\rr{h}))$ is weakly contractible (\ie all of its homotopy groups are trivial). 
As a consequence, one has that 
$[X,B(^{\rm w}\bb{U}(\rr{h}))]$ reduces to a single point, and in turn
\[
{\rm Hilb}_{\rr{h}}(X)\;=\;[\bb{H}_0]
\]
 where on the right-hand side one has the class of the trivial Hilbert bundle.

\subsection{From group action to bundles}\label{app:G-bund}
The construction described in this section is inspired by \cite[Section 1.6]{atiyha-67}. 

\medskip

Let $X$ be a locally compact Hausdorff space and $\Gamma$ a discrete group.
One says that $X$ is a $\Gamma$-space if $\Gamma$ acts on $X$ and the map 
\[
\Gamma\times X\;\ni(\gamma,x)\longmapsto \gamma\cdot x\;\in\;X
\]
is continuous where $\gamma\cdot x$ denotes the action of the element $\gamma\in\Gamma$ on $x\in X$. The action of $\Gamma$ is \emph{free} if $\gamma\cdot x=x$ for some $x\in X$ implies that $\gamma={\imath}$ where $\imath$ denotes the identity of the group.  Since $\Gamma$ is discrete, this condition guarantees that the orbit space $X_\Gamma:= X/\Gamma$ is Hausdorff in the quotient topology.  
\medskip

Let $\Gamma\ni \gamma\mapsto U_\gamma\in\bb{U}(\rr{h})$ be a unitary representation of $\Gamma$ on $\rr{h}$. The {trivial} Hilbert bundle $\bb{H}_0:=X\times \rr{h}\to X$ inherits a $\Gamma$-action given by $\gamma:(x,f)\mapsto(\gamma\cdot x,U_\gamma f)$, for every $(x,f)\in \bb{H}_0$. Such an action commutes with the bundle projection, and is evidently free. Consider the quotient space $\bb{H}_\Gamma:=\bb{H}_0/\Gamma$ endowed with the quotient topology and the projection $\pi:\bb{H}_\Gamma\to X_\Gamma$ given by $\pi:[(x,f)]\mapsto[x]$. This map is well defined since it doesn't depend on the choice of the representative. The bundles $\bb{H}_\Gamma\to X_\Gamma$ and $\bb{H}_0\to X$ are locally isomorphic, and since $\bb{H}_0\to X$ is (locally) trivial, it follows that also 
$\bb{H}_\Gamma\to X_\Gamma$ is locally trivial. It turns out that $\bb{H}_\Gamma\to X_\Gamma$ is a Hilbert bundle with typical fiber $\rr{h}$. 
In view of the infinite-dimensionality of the typical fiber, one has that 
$\bb{H}_\Gamma\simeq X_\Gamma\times \rr{h}$, as explained at the end of Appendix \ref{app:hilb_bun}.

\medskip

The same construction can be lifted to the case of Grassmann bundles. First of all, $\Gamma$ acts on the bounded operators $\bb{B}(\rr{h})$ by conjugation, \ie, $\lambda_\gamma(A) := U_\gamma^*AU_\gamma$ for all $A\in \bb{B}(\rr{h})$ and $\gamma\in \Gamma$. Evidently, this action restricts to any (weak) Hilbert Grassmannian 
$${^{\rm w}}\bb{G}_k(\rr{h}) \;:=\; \left\{P\in\bb{P}(\rr{h})\;|\; {\rm Tr}_{\rr{h}}(P)=k \right\}\;,$$ 
since $\lambda_\gamma(\bb{G}_k(\rr{h}))=\bb{G}_k(\rr{h})$ for every $\gamma\in \Gamma$. The superscript ${\rm w}$ means the space is equipped with the weak topology.
Consider the trivial Grassmann ($k$-plane) bundle $X \times {^{\rm w}}\bb{G}_k(\rr{h})\to X$
and the $\Gamma$-action $\gamma:(x,P)\mapsto(\gamma\cdot x,\lambda_\gamma(P))$. Let 
$\bb{G}_k^\Gamma:=(X \times {^{\rm w}}\bb{G}_k(\rr{h}))/\Gamma$ endowed with the quotient topology and the projection $\bb{G}_k^\Gamma\to X_\Gamma$ given by
$\pi:[(x,P)]\mapsto[x]$. Similarly to the previous case, it turns out that $\bb{G}_k^\Gamma\to X_\Gamma$ is a Grassmann bundle with typical fiber ${^{\rm w}}\bb{G}_k(\rr{h})$.
Moreover, one can check that $\bb{G}_k^\Gamma$ is the Grassmann bundle subordinated to $\bb{H}_\Gamma$, \ie $\bb{G}_k^\Gamma=\rr{Gr}_k(\bb{H}_\Gamma)$. Again, in view of the triviality of the underlying bundle one gets that $\bb{G}_k^\Gamma\simeq X_\Gamma\times {^{\rm w}}\bb{G}_k(\rr{h})$. 
In \cite{denittis-gomi-rendel-25}, we will study the spaces ${^{\rm w}}\bb{G}_k(\rr{h})$ and their use as classifying spaces.

\section{An overview on the Bohr compactification}\label{appendix:Bohr}
In this section we will review some basic facts about the Bohr compactification, and prove some crucial results used in the main part of this work. The presentation is based mainly on \cite[Section 1.8]{rudin-62},  \cite[Section 3.4]{dikranjan-prodanov-stoyanov-90}, and \cite{hewitt-53}. For notational simplicity, we will identify any locally compact abelian group $\n{G}$ with its \emph{Pontryagin bidual}, \ie $\n{G} \equiv \doublehat{\n{G}}$.

\subsection{General facts}\label{app:gen_fct}

As it is well known, the Pontryagin dual of a discrete abelian group is a compact group. Let $\n{G}$ be any locally compact abelian (LCA) group, with Pontryagin dual $\widehat{\n{G}}$. Denote by $\widehat{\n{G}}_d$ the group $\widehat{\n{G}}$ equipped with the \emph{discrete topology}.
Since  $\widehat{\n{G}}_d$ is a discrete group, its Pontryagin dual 
\[
\rr{b}(\n{G})\;:=\;\widehat{\widehat{\n{G}}_d}
\] 
is a compact Hausdorff group, called the \emph{Bohr compactification} of $\n{G}$. Observe that, when $\n{G}$ is compact, one has that $\rr{b}(\n{G}) = \n{G}$.

\medskip

Note that, in the spirit of the Pontryagin duality,
$\rr{b}(\n{G})$ may be interpreted as the set of all \virg{continuous}
characters of $\widehat{\n{G}}_d$, which are in fact all possible characters in view of the discreteness of the topology. 
In this sense $\rr{b}(\n{G})$ contains $\n{G}$, which can be interpreted as the group of continuous characters of $\widehat{\n{G}}$ (with respect to its original topology).
Let us make this more precise. Let $\rr{i}: \n{G} \to \rr{b}(\n{G})$ be the map 
\[
\big(\rr{i}(g) \big) (\zeta) \;:=\; \zeta \left(  g \right) \,,\quad \forall \;g\in \n{G},\; \quad \forall \;\zeta\in \widehat{\n{G}}\;.
\]
We denote by $\n{G}_\rr{b}:=\rr{i}(\n{G})$ its image inside $\rr{b}(\n{G})$, and define $\rr{i}_{\rr{b}}:\n{G} \to \n{G}_\rr{b}$ to be the range restriction of $\rr{i}$.

\begin{theorem}[{\cite[Theorem 1.8.2]{rudin-62}}]\label{thm:bohr_comp_map}
    The map $\rr{i}_{\rr{b}}:\n{G} \to \n{G}_\rr{b}$ is a continuous group isomorphism, and $\n{G}_\rr{b}$ is a dense subgroup of $\rr{b}(\n{G})$.
\end{theorem}

\medskip

This result allows us to interpret $\n{G}$ as a dense subgroup of its Bohr compactification $\n{G}_\rr{b}\subset\rr{b}(\n{G})$, albeit with a different topology. We will refer to $\n{G}_\rr{b}$ as $\n{G}$ endowed with the \emph{Bohr topology}.
It is important to remark that $\rr{i}_{\rr{b}}$ is \emph{not} an homeomorphism, and $\n{G}_\rr{b}$ is not a locally compact subspace of $\rr{b}(\n{G})$ unless $\n{G}$ is compact.

\medskip

    A basis for the topology of $\rr{b}(\n{G})$ is given by the sets
    \[
    {\s{O}}_{\{\zeta_1,\dots,\zeta_m\},\varepsilon}(g) \;:=\; \left\{ g'\in \rr{b}(\n{G}) \; \big| \; |g(\zeta_j) - g'(\zeta_j)|<\varepsilon , \, j=1\dots,m \right\}, 
    \]
    where $g\in \rr{b}(\n{G})$, $\zeta_1,\dots,\zeta_m \in \widehat{\n{G}}$ and $\varepsilon>0$.
    This topology is not metrizable, but net convergence can be easily characterized. Let $\{\lambda_i\}_{i\in\bb{I}} \subset \rr{b}(\n{G})$ be a net, and $\lambda\in \rr{b}(\n{G})$. Then,
    \[
    \lambda_i\to \lambda \; \iff\; \lambda_i(\zeta) \to \lambda(\zeta) \,, \quad \forall\; \zeta \in \widehat{\n{G}}.
    \]
Of course, when restricting to $\n{G}_\rr{b}$  the condition of net convergence reads $g_i\to g \iff \zeta(g_i)\to \zeta(g)$, for all $\zeta\in \widehat{\n{G}}$. Indeed, the Bohr topology is nothing more that the weak topology in $\n{G}$. 
In view of this, it turns out that the Bohr topology is strictly weaker than the original topology of $\n{G}$. This fact is, however, not so dramatic. In fact, since $\n{G}_\rr{b}$ is not locally compact, the basis of the topology is not forced to have any pre-compact set. Then, one can  hope that at least compact sets are \virg{small enough} so that they behave like in the original topology. This happens to be the case.
\begin{theorem}[{\cite[Theorem 3.4.3]{dikranjan-prodanov-stoyanov-90}}]\label{thm:compacts_bohr}
    Let $K$ be a compact subset of $\n{G}$ and consider its image $K_\rr{b} := \rr{i}_\rr{b}(K)$ in $\rr{b}(\n{G})$. Then $K$ endowed with the topology inherited from $\n{G}$ and $K_\rr{b}$ with the 
    Bohr topology are homeomorphic, \ie
     $K\simeq K_\rr{b}$. Moreover  every compact set of $\n{G}_\rr{b}$ is of this form.
\end{theorem}

\medskip
One implication of the result above is quite simple.
Since $\rr{i}_\rr{b}$ is continuous and bijective,
$K$ is compact and $K_\rr{b}$ is  Hausdorff then  
 the homeomorphism $K\simeq K_\rr{b}$ is immediate. Therefore,  every compact set in $\n{G}$ is also compact in $\n{G}_\rr{b}$. 
The proof that every compact set in $\n{G}_\rr{b}$ is also compact in $\n{G}$ is more technical and  we refer to cited paper for the details.

\medskip

One other relevant property of the Bohr topology is the following:
\begin{theorem}[\cite{reid-67}]
    $\n{G}_\rr{b}$ is sequentially closed, and $\rr{i}_\rr{b}: \n{G} \to \n{G}_\rr{b}$ is a sequential homeomorphism (\ie its inverse is sequentially continuous).
\end{theorem}

Finally, we present an important characterization of the Bohr compactification, often presented as its definition.

\begin{theorem}[{\cite{holm-64}}]\label{thm:univ_prop_bohr}
    The Bohr compactification of $\n{G}$ satisfies the following universal properties:
    \begin{enumerate}
        \item[\emph{(i)}] For any continuous homomorphism $\phi:\n{G}\to \n{K}$, with $\n{K}$ a compact Hausdorff abelian group, there exists a continuous homomorphism $\widehat{\phi}: \rr{b}(\n{G}) \to \n{K}$ such that $${\phi}\; =\; \widehat{\phi}\circ \rr{i}\;.$$
        \item[\emph{(ii)}] If in addition ${\n{G}'}$ is another compact group such that $\rr{i}':\n{G}\to {\n{G}'}$ is a continuous  homomorphism with dense image and the pair $(\rr{i}', {\n{G}'})$ meets property (i), then there is an isomorphism of topological groups
 $\rr{b}(\n{G})\simeq\n{G}'$.
    \end{enumerate}
\end{theorem}

\medskip

The result above states that the map $\rr{i}$ is \emph{universal} with respect to continuous homomorphisms into compact groups, and that the pair $(\rr{i},\rr{b}(G))$ is characterized by this \emph{universal} property up to isomorphisms.

\subsection{Relation with the almost periodic functions}\label{sec_AP}
In this section we provide a  brief presentation of the relationship between the Bohr compactification and the almost periodic functions of a group.  Let $\n{G}$ be LCA and interpret $\widehat{\n{G}}$ as a subset of $C_{\rm b}(\n{G})$, where the latter is the $C^*$-algebra of complex-valued, bounded continuous functions.
The functions
\[
TP(\n{G}) \;:=\; {\rm span}\,  (\widehat{\n{G}} ) 
\]
of the form $f = \sum_{j=1}^k c_j \zeta_j$ for $c_j\in \C$ and  $\zeta_j\in\widehat{\n{G}}$, are called the \textit{trigonometric polynomials} of $\n{G}$. They form a self-adjoint subalgebra of $C_{\rm b}(\n{G})$. Its closure, denoted by
\[
AP(\n{G}) \;:=\; \overline{TP(\n{G})}^{\;\|\cdot\|_\infty}
\]
is the set of \textit{almost periodic functions} of $\n{G}$. This is a unital $C^*$-subalgebra of $C_b(\n{G})$, hence also abelian. Thus, by the Gelfand isomorphism, there exists a compact Hausdorff space $X$ such that $AP(\n{G}) \simeq C(X)$. It turns out that $X$ is exactly $\rr{b}(\n{G})$.
\begin{proposition}\label{prop:isom_CBohr_AP}
    The following $*$-isomorphisms hold true
    \[
    AP(\n{G})\; \simeq\; C^*(\widehat{\n{G}}_d)\; \simeq\; C(\rr{b}(\n{G}))
    \]
     Moreover, the correspondence $C(\rr{b}(\n{G})) \to AP(\n{G})$ is given by $f\mapsto f\circ\rr{i}$, interpreted as restricting the function $f$ to $\n{G}$.
\end{proposition}

\medskip

The result above is classical and can be found in various sources like \cite{hewitt-53,bottcher-karlovich-spitkovsky-02,binz-honegger-rieckers-04}. It is worth it to point out that the identification $AP(\n{G})\simeq C(\rr{b}(\n{G}))$ can be understood in the following sense: every function in $AP(\n{G})$ admits a (unique) extension to a function in $C(\rr{b}(\n{G}))$ and conversely, the restriction of every function in $C(\rr{b}(\n{G}))$ to $\R^d$ is a function in $AP(\n{G})$ (\cf \cite[Theorem 7.4]{bottcher-karlovich-spitkovsky-02}).

\subsection{The case of  \texorpdfstring{$\n{G} = \R^d$}{TEXT}}
 Since $\widehat{\R^d} = \R^d$, the Bohr compactification of $\R^d$ is given by 
\[
\rr{b}(\R^d)\;: =\; \widehat{\R^d_d}\;.
\]  
The following result is of a crucial relevance. 
\begin{theorem}[{\cite[Theorem 1.4]{hewitt-53}}]\label{thero:cont_car}
    The Bohr compactification $\rr{b}(\R^d)$ corresponds to all possible characters of $\R^d$. The continuous characters of $\R^d$ are exactly the elements of $\R^d_\rr{b}\subset \rr{b}(\R^d)$.
\end{theorem}

\medskip

Let us recall that any continuous character of $\R^d$ has the form 
$x\mapsto \expo{\ii \beta \cdot x}$ for some $\beta\in\R^d$.
We denote the characters of $\rr{b}(\R^d)$ (corresponding to points of $\R^d_d$) by $\xi_\beta$ and are indeed labeled by $\beta \in \R^d$ (as a set). Of course, in view of the result above and biduality, $\xi_\beta(x) = \expo{\ii \beta \cdot x}$ for any $x\in\R^d_\rr{b}\subset \rr{b}(\R^d)$. Net convergence is given in this case by 
\[
\lambda_i \to \lambda \; \iff \; \xi_\beta(\lambda_i) \to \xi_\beta(\lambda) \,, \quad \forall\; \beta \in\R^d
\]
where $\{\lambda_i\}_{i\in\bb{I}} \subset \rr{b}(\R^d)$ is any net converging to $\lambda\in\rr{b}(\R^d)$. This implies that 
$x_i\to x$ in $\R^d_\rr{b}$ if and only if $\expo{\ii x_i \cdot \beta} \to \expo{\ii x \cdot \beta}$ for every $\beta \in\R^d$.

\medskip

The connection properties of $\rr{b}(\R^d)$ are described in the next result.

\begin{proposition}\label{prop:conex_bohr}
    The space $\rr{b}(\R^d)$ 
    is connected but not path-connected. 
    \end{proposition}
\proof  
First we prove the connectivity. By \cite[Theorem 46]{pontryagin-86}, if a compact abelian group $\n{G}$ has a torsion-free Pontryagin dual, then it is connected. In our case, $\rr{b}(\R^d)$ is compact and abelian, and its dual $\R^d_d$ is clearly torsion free, hence the claim follows.

Now we prove the path-disconnectedness. Dixmier proved in \cite{dixmier-57} that a compact abelian group $\n{G}$ is path-connected if and only if $\operatorname{Ext}^1(\widehat{\n{G}},\Z) = \{0\}$ is the trivial group, where $\widehat{\n{G}}$ is the Pontryagin dual of $\n{G}$, and $\operatorname{Ext}^1(\cdot,\Z)$ is the first right derived functor of $\operatorname{Hom}(\cdot,\Z)$ on the category of abelian groups. Therefore, it suffices to show that $\operatorname{Ext}^1(\R^d_d,\Z) \neq \{0\}$. To this purpose, let $\rr{B}$ be a Hamel basis of $\R^d$ over $\Q$. Then 
\[
\R^d_d \simeq \bigoplus_{\rr{B}} \Q_d.
\]
Where $\Q_d$ are the rationals with the discrete topology. Therefore,
\begin{align*}
    \operatorname{Ext}^1\left(\R^d_d,\Z\right) \;\simeq\; \operatorname{Ext}^1\left(\bigoplus_{\rr{B}} \Q_d,\Z\right) \;\simeq\; \prod_{\rr{B}} \operatorname{Ext}^1(\Q_d,\Z)
    \simeq \;\prod_{\rr{B}} \R_d \neq \{0\}\,.
\end{align*}
Where in the second isomorphism we used \cite[Chapter III, Lemma 4.1]{hilton-stammbach-12}
and in the third we used the fact that $\operatorname{Ext}^1(\Q_d,\Z)\simeq \R_d$ \cite[Chapter III, exercise 6.2]{hilton-stammbach-12}.
\qed

\subsection{Measure theory}\label{app:mas_thr}

In the following we will present some relevant facts about the measures of $\rr{b}(\R^d)$ and the states of $C(\rr{b}(\R^d))$. Our main reference will be \cite{hewitt-53}. For this, we first study the measures of the Bohr compactification and the Bohr topology. For practical reasons we consider $\R^d_\rr{b}$ and $\R^d$ to be the same set with different topologies.

\medskip

First of all, note that the Borel $\sigma$-algebra $\s{B}(\R^d)$ of $\R^d$ contains the Borel $\sigma$-algebra $\s{B}(\R^d_\rr{b})$ of $\R^d_\rr{b}$ in a natural sense, as the Bohr topology is weaker than the usual topology. As such, any measure defined on $\s{B}(\R^d)$ defines a measure in $\s{B}(\R^d_\rr{b})$ by restriction provided that this restriction preserves regularity. Let us denotes with 
$\s{M}_{+,1}(\R^d)$ and $\s{M}_{+,1}(\R^d_\rr{b})$ the spaces of the normalized positive regular measures defined on  $\s{B}(\R^d)$ and  $\s{B}(\R^d_\rr{b})$, respectively. Both spaces are meant endowed with respect to the  $\ast$-weak topology.
\begin{lemma}\label{lemma:regular_measure_embed}
    The restriction of a regular measure $\mu\in \s{M}_{+,1}(\R^d)$ to $\s{B}(\R^d_\rr{b})$ defines a regular measure $\widetilde{\mu}\in\s{M}_{+,1}(\R^d_\rr{b})$. Moreover, the restriction map defines an homeomorphism $I_\s{M}:\s{M}_{+,1}(\R^d)\rightarrow \s{M}_{+,1}(\R^d_\rr{b})$.
\end{lemma}
\proof 
The inner regularity of $\widetilde{\mu}$ is immediate from Theorem \ref{thm:compacts_bohr}, since for any $A \in \s{B}(\R^d_\rr{b})$,
\begin{align*}
    \widetilde{\mu}(A)\; =\; \mu(A) &\;=\; \sup \, \left\{ \mu(K) \;\left|\; K\subseteq A \text{ compact in } \R^d \right\}\right. \\ 
    &\;=\; \sup \, \left\{ \widetilde{\mu}(K) \;\left|\; K\subseteq A \text{ compact in } \R^d_\rr{b} \right\}\right.\,. 
\end{align*}
Then, since $\widetilde{\mu}$ is finite and $\R^d_\rr{b}$ is Hausdorff, it is also regular, as a consequence of \cite[Proposition 1]{gruenhage-pfeffer-78} (do note that our notion of (inner) regularity is what is called \emph{compact} (inner) regularity in this paper).  Suppose $\mu_1,\mu_2\in\s{M}_{+,1}(\R^d)$ are two measures with the same restriction $\widetilde{\mu}$. Then, for any $A\in \s{B}(\R^d_\rr{b})$,
\begin{align*}
    \mu_1(A) &\;=\; \sup \, \left\{ \mu_1(K) \;\left|\; K\subseteq A \text{ compact in } \R^d \right\}\right.\\
    &\;=\; \sup \, \left\{ \widetilde{\mu}(K) \;\left|\; K\subseteq A \text{ compact in } \R^d_\rr{b} \right\}\right.\\
    &\;=\; \sup \, \left\{ \mu_2(K) \;\left|\; K\subseteq A \text{ compact in } \R^d \right\}\right.\;=\; \mu_2(A).
\end{align*}
Hence $I_\s{M}$ is injective. {To prove that it is surjective, let $\widetilde{\mu}\in \s{M}_{+,1}(\R^d_\rr{b})$, and define
\begin{align*}
    \mu_i(A) &\;:=\; \sup\,\big\{ \widetilde{\mu}(K) \;\big|\; K\subseteq A \text{ compact in }\R^d_\rr{b} \big\} \\
    \mu_e(A) &\;:=\; \inf\,\left\{ \widetilde{\mu}(\s{O}) \;\left|\; \s{O}\supseteq A \text{ open in }\R^d_\rr{b} \right.\right\}
\end{align*}
for any $A\subseteq\R^d$. It can be easily checked that $\mu_i$ is an inner measure and $\mu_e$ an exterior (or outer) measure. Moreover,   both coincide with $\widetilde{\mu}$ on $\s{B}(\R^d_\rr{b})$ and in particular in the compact sets of $\R^d$. Therefore, the $\sigma$-algebra $\Sigma$ on which they coincide contains the closed sets of $\R^d$, as any closed set is a countable union of compact sets. We conclude that $\Sigma$ contains $\s{B}(\R^d)$, and therefore $\mu = \mu_i|_{\s{B}(\R^d)} = \mu_e|_{\s{B}(\R^d)}$ defines a positive probability measure, regular by construction, such that its restriction $I_\s{M}(\mu)$ is $\widetilde{\mu}$.}
{Now, to prove continuity, let $\mu\in\s{M}_{+,1}(\R^d)$, and $\{\mu_i\}_{i\in\bb{I}}\subset \s{M}_{+,1}(\R^d)$ a net such that $\mu_i\to\mu$. Denote $\widetilde{\mu}:= I_\s{M}(\mu)$ and $\widetilde{\mu}_i := I_\s{M}(\mu_i)$. First note that if $f:\R^d_\rr{b}\to\C$ is measurable, then for any Borel subset $E$ of $\C$, $f^{-1}(E) \in \s{B}(\R^d_\rr{b})\subset \s{B}(\R^d)$, and thus $f$ is also measurable as a function $\R^d\to\C$. Let $f\in C_0(\R^d_\rr{b})\subseteq C_0(\R^d)$. Then, approximating $f$ pointwise by simple functions in $L^1(\R^d_\rr{b},\widetilde{\mu})$ and $L^1(\R^d_\rr{b},\widetilde{\mu}_i)$ and applying the Dominated Convergence Theorem,
\begin{align*}
    \int_{\R^d_\rr{b}} f \dd\widetilde{\mu} - \int_{\R^d_\rr{b}} f \dd\widetilde{\mu_i} &\;=\; \lim_{N\to\infty} \left( \sum_{j=1}^N a_{N,j}\, \widetilde{\mu}(E_{N,j}) - \sum_{j=1}^N a^i_{N,j} \,\widetilde{\mu}(E^i_{N,j}) \right) \\
    &\;=\; \lim_{N\to\infty} \left( \sum_{j=1}^N a_{N,j}\, {\mu}(E_{N,j}) - \sum_{j=1}^N a^i_{N,j} \,{\mu}(E^i_{N,j}) \right) \\
    &\;=\; \int_{\R^d} f \dd{\mu} - \int_{\R^d} f \dd{\mu_i} \to 0\;.
\end{align*}
Therefore, $I_\s{M}(\mu_i)\to I_{\s{M}}(\mu)$, and $I_\s{M}$ is continuous.
}
Hence, since $I_\s{M}$ is bijective, continuous, and has compact domain, it is an homeomorphism.
\qed

\medskip

\begin{remark}
    The fact that $\s{M}_{+,1}(\R^d)$ and $\s{M}_{+,1}(\R^d_\rr{b})$ are homeomorphic can also be deduced from the observation that $C_0(\R^d_\rr{b}) \simeq C_0(\R^d)$, since both topologies share exactly the same compact sets. \hfill $\blacktriangleleft$
\end{remark}

\medskip

The main message of Lemma \ref{lemma:regular_measure_embed} is that every measure in $\s{M}_{+,1}(\R^d_\rr{b})$ has a unique regular extension on $\s{M}_{+,1}(\R^d)$. By combining this fact with  \cite[Theorem 4.1]{hewitt-53} one obtains the following result.

\begin{corollary}\label{corol:restr_measu}
    Let $\mu\in\s{M}_{+,1}(\rr{b}(\R^d))$ and denote with $\xi_\beta$ the character of $\rr{b}(\R^d)$ labelled by $\beta\in\R^d$ (as a set). Consider the function $f_\mu:\R^d\to\C$ given by

    \[
   f_\mu (\beta)\; :=\; \int_{\rr{b}(\R^d)}\dd \mu(\lambda)\; \xi_{-\beta}(\lambda)  \;.
    \]
Then $f_\mu$  is continuous with respect to the standard topology of $\R^d$ if and only if the measure $\mu$ is supported in $\R^d_\rr{b}$. In such case one has 
    \[
     f_\mu (\beta)\; =\; \int_{\R^d}\dd \widehat{\mu}(\lambda)\; \expo{-\ii \beta\cdot\lambda}  
    \]
    where $\widehat{\mu}$ is the unique regular extension of $\mu$.
\end{corollary}

\end{document}